\documentclass[aps,pre,preprint,groupedaddress,amssymb]{revtex4-2}

\usepackage{graphicx} 
\usepackage{dcolumn}
\usepackage{bm}
\usepackage{amsmath}
\usepackage{epsfig}
\usepackage{color,hyperref} 
\usepackage{amssymb}
\usepackage{appendix}

\begin{document}

\title{Absolute area under a scaled Brownian motion with resetting}

\author{Vicen\c c M\'endez}
\email{vicenc.mendez@uab.cat}
\author{Javier Crist\'in}

\affiliation{Grup de F\'{\i}sica Estad\'{\i}stica, Departament de F\'{\i}sica. Facultat de Ci\`{e}ncies, Universitat Aut\`{o}noma de Barcelona, 08193 Barcelona, Spain}

\begin{abstract}
We investigate the statistical properties and ergodic behavior of the absolute area below the trajectory of a particle subject to stochastic resetting and moving according to a scaled Brownian motion with exponent $\beta$. We compare the contributions to the absolute area from the waiting times between resets and the backward recurrence time. Using the renewal formalism we derive the lower-order moments of the absolute area under both Poissonian and heavy-tailed power-law resetting waiting time distributions with exponent $\alpha$. For Poissonian resetting, stochastic interruption suppresses fluctuations, driving the area to be ergodic, regardless of the value of $\beta$. In contrast, power-law resetting yields a rich spectrum of ergodic regimes. For $0 < \alpha < 1$, the system displays weak ergodicity breaking and the maximum contribution dominates the area accumulated during the waiting times. For $1+\beta < \alpha < 2$, we unveil an ergodic transition exclusive to the subdiffusive regime  at $\alpha=1+\beta$, where ergodicity is restored. These results clarify the subtle interplay between anomalous diffusion and power-law waiting times between resettings. Our results are tested by numerical simulations exhibiting an excellent agreement.
\end{abstract}

\maketitle

\section{Introduction}

Over the past decade, stochastic resetting has emerged as a paradigmatic framework in non-equilibrium statistical physics \cite{Evans2011,Evans2020}. Stochastic interruption fundamentally alters transport properties, prevents unbounded spatial spreading, and optimizes search strategies across physical, biological, and algorithmic systems \cite{Evans2020,Gu22}. While early studies focused primarily on simple Markovian Brownian motion \cite{Evans2011,Evans2020}, significant efforts have expanded this paradigm to non-Markovian dynamics and complex media, including continuous-time random walks (CTRW), fractional Brownian motion, and scaled Brownian motion (SBM) \cite{BoSo20,EuMe16,BoChe19,MeMaCa22,MeSa21}. Scaled Brownian motion, in particular, offers a tractable model for anomalous diffusion driven by an explicit, deterministic time-dependent diffusion coefficient $D(t) \sim t^{\beta-1}$, capturing subdiffusive ($\beta < 1$) or superdiffusive ($\beta > 1$) spatial kinetics common in crowded cellular media and active transport \cite{Lim2002,Jeon2014}. While spatial probability densities under resetting are now well understood, time-integrated observables provide a much deeper characterization of trajectory dynamics, residence statistics, and path-to-path heterogeneity \cite{Kac1949,Majumdar2005}. 
Among these observables, the area and the absolute area under the trajectory as a function of observation time have been widely studied, particularly for the standard Brownian motion \cite{Sh82,PeWe96,Ta93,MaBr02}. 
The absolute area was first studied in the context of economics \cite{Ci75} and later extensively in probability \cite{Sh82,Ta93}. The same functional was also studied in the context of electron-electron and phase coherence in one dimensional weakly disordered quantum wire \cite{Al82}. The area has also been studied when the random walk is a Brownian motion under a potential \cite{Sm24} or under resetting \cite{Ho19}.

The ergodic properties of stochastic processes under resetting depend sensitively on the nature of the resetting time distribution $\varphi(t)$ \cite{BaFlMe23}. Under standard Poissonian resetting with a constant rate $r$, the system possesses a characteristic time scale $\langle \tau \rangle = 1/r$ that repeatedly erases spatial correlations, effectively suppressing long-time trajectory fluctuations and restoring ergodicity \cite{Sto22,WaCh22}. However, in complex or disordered environments, resetting events may follow non-Poissonian, heavy-tailed waiting-time distributions $\varphi(t) \sim t^{-1-\alpha}$, where rare, exceptionally long renewal intervals dominate the time evolution \cite{NaGu16,ChSo18,MaCaMe19}. In such heavy-tailed regimes, resetting competes directly with the intrinsic temporal scaling of the underlying random walk, giving rise to weak ergodicity breaking (WEB), non-stationary statistics \cite{BaFlMe23}.

Despite the extensive literature on resetting and stochastic functionals, a unified theoretical understanding of the absolute area under a non-Brownian trajectory subject to resetting remains incomplete. Specifically, how the anomalous exponent $\beta$ of scaled Brownian motion interacts with the heavy-tailed resetting exponent $\alpha$ to dictate the statistical properties of the absolute area remains an open question. In this work, we address this gap by computing lower-order moments of the absolute area and analyzing the asymptotic behavior of the ergodicity breaking parameter for both Poissonian and power-law resetting rules. We explore in detail the statistical properties of the contributions to the absolute area from the backward time and the waiting times between resets. This analysis helps to understand the different ergodic regimes and which is the main contribution depending on the values of the $\alpha$ and $\beta$ exponents.

The paper is organized as follows. In Sec.~\ref{sec:framework}, we construct the general renewal formalism for stochastic functionals under resetting. In Sec.~\ref{sec:ergodic_properties}, we define the ergodicity breaking parameter and establish its connection to cumulative path observables. In Sec.~\ref{sec:absolute_area}, we evaluate the absolute area functional for scaled Brownian motion under Poissonian and heavy-tailed power-law resetting, identifying a novel ergodic transition in the subdiffusive regime. In Secs.~\ref{sec:backward} and \ref{sec:waiting_time}, we analyze separately the contributions from the backward recurrence time and during the waiting periods between resets, analyzing the maximum area via extreme value statistics. Finally, Secs.~\ref{sec:discussion} and \ref{sec:conclusions} discuss the physical implications of our findings and offer concluding remarks.

\section{Stochastic functionals under resetting}
\label{sec:framework}
Let $Z_r(t)$ be the stochastic functional of the trajectory $\{x(t'); 0\leq t'\leq t\}$ of a particle forced to reset at random times to its initial position $x(0)=0$. By definition,
\begin{eqnarray}
Z_r(t)=\int_{0}^{t}U[x(t')]_rdt'
    \label{Z}
\end{eqnarray}
where the subindex $r$ indicates that the random walk is under the effect of resetting. $U[x(t)]_r$ is a function of the trajectory.

The random walk process under resetting is as follows. The walker starts moving at $x(0)=0$, a positive resetting time $\tau_1$ is drawn from the PDF $\varphi (\tau)$ and the particle performs a free random walk in the interval of time $(0,\tau_1)$, finally reaching some random position $x(\tau_1)$. Then the particle's position is reset instantaneously to $x=0$, and the process is then renewed, namely, a second resetting
time $\tau_2$ is drawn also from $\varphi (\tau)$, etc. When this process is continued we get the sequence of independent identically distributed (IID) random variables, $\{\tau_1, \tau_2,...\}$, i.e., the waiting times between resetting events, which are needed to construct the path of the particle. 
Let $P_N(Z_r,t)dtdZ_r$ be the probability that the $N$-th resetting event takes place in the time interval $(t,t+dt)$ when the value of the functional is within $(Z_r,Z_r+dZ_r)$. This probability obeys the conservation equation for the probability
\begin{eqnarray}
 P_{N+1}(Z_{r},t)=\int_{0}^{Z_{r}}dZ_{r}'\int_{0}^{t}d\tau P_{N}(Z_{r}-Z_{r}',t-\tau)\phi(Z_{r}',\tau),
 \label{eb1}
\end{eqnarray}
where $\phi (Z_{r},\tau)$ is the joint PDF of the waiting  period between consecutive resets and the functional. Eq. \eqref{eb1} describes the evolution of the process: the functional takes the value $Z_r$ at time $t$ when the previous reset took place at time $t-\tau$, the previous value of the functional was $Z_r-Z_r'$. Consider that the last reset event is at time $t-B$ and $B$ is called the backward recurrence time. According to $t=\tau_1+\cdots+\tau_N+B$ the resetting process is defined with a sequence of waiting times (time intervals) between resettings. Within each $\tau_i$ with $i=1,\cdots,N$ we have a contribution $Z_{0}^{(i)}=\int_{0}^{\tau_{i}}U[x(t')]dt'$ to $Z_r$. $U[x(t)]$ is a function of the trajectory of the random walk without resettings. Clearly, the functional under resetting can be written as the sum of contributions from the resettings plus the contribution $Z_{\textrm{back}}$ from the backward time
\begin{eqnarray}
    Z_{r}=\sum_{i=1}^{N}Z_{0}^{(i)}+Z_{\textrm{back}}.
    \label{contrib}
\end{eqnarray}
The sets $\{Z_{0}^{(1)},\cdots,Z_{0}^{(N)}\}$ and $\{\tau_{1},\cdots,\tau_{N}\}$ are separately composed from 
IID random variables. However, $Z_0^{(i)}$ and $\tau_i$ are mutually dependent. The longer is $\tau_i$ the longer $Z_{0}^{(i)}$ is. Their joint PDF is $\phi (Z_0,\tau)=\varphi(\tau)P(Z_0|\tau)$. $P(Z_0|\tau)$ is the PDF of the functional without resettings for a fixed measurement time $\tau$. We want to solve \eqref{eb1} recursively. To this end we introduce first the double Laplace transform of $P_{N}(Z_{r},t)$ with respect to $Z_{r}$ and $t$: 
\begin{eqnarray}
    \tilde{Q}_N(p,s)_r&=&\mathcal{L}_{Z_r\to p}\mathcal{L}_{t\to s}[P_N(Z_r,t)]=\int_{0}^{\infty}dte^{-st}\int_{0}^{\infty}dZ_re^{-pZ_r}P_N(Z_r,t).
\end{eqnarray}
Performing the double Laplace transform to Eq. \eqref{eb1} one has $\tilde{Q}_{N+1}(p,s)_r=\tilde{Q}_{N}(p,s)_r\tilde{\phi}(p,s)$ which can be iteratively solved using the initial condition $P_0(Z_r,t)=\delta (Z_r)\delta (t)$, which means that the resetting process starts at $t=0$ with $Z_t=0$. The solution is $\tilde{Q}_{N}(p,s)_r=\tilde{\phi}(p,s)^N$. On the other hand, according to  \eqref{contrib} the PDF of the functional under resetting is the convolution of the PDF corresponding to the contribution of the resets and the PDF corresponding to the backward time
\begin{eqnarray}
    P(Z_{r},t)=\int_{0}^{Z_{r}}dZ_{\textrm{back}}\int_{0}^{t}dB\sum_{N=0}^{\infty}P_{N}(Z_{r}-Z_{\textrm{back}},t-B)\Phi(Z_{\textrm{back}},B),
    \label{pzr}
\end{eqnarray}
where we have summed over the number of resettings $N$ within $[0,t]$. $\Phi(Z_{\textrm{back}},B)$ is the joint PDF of $Z_{\textrm{back}}$ during the backward time so that $\Phi(Z_{\textrm{back}},B)=\varphi ^* (B)P(Z_{\textrm{back}}|B)$ where $\varphi ^* (B)$ is the probability of no reset within the backward time, i.e.,
$$
\varphi ^* (\tau)=1-\int_0^\tau \varphi (\tau')d\tau'=\int_\tau^\infty \varphi (\tau')d\tau'.
$$ 
Computing the double Laplace transform to Eq. \eqref{pzr} and summing the geometric series we have
\begin{eqnarray}
    \tilde{Q}(p,s)_{r}&=&\int_{0}^{\infty}dte^{-st}\int_{0}^{\infty}dZ_{r}e^{-pZ_{r}}P(Z_{r},t)=\sum_{N=0}^{\infty}\tilde{Q}_{N}(p,s)\tilde{\Phi}(p,s)=\frac{\tilde{\Phi}(p,s)}{1-\tilde{\phi}(p,s)}\nonumber\\
    &=&\frac{\mathcal{L}_{t\to s}\left[\Phi(p,t)\right]}{1-\mathcal{L}_{t\to s}\left[\phi(p,t)\right]}=\frac{\mathcal{L}_{t\to s}\left[\varphi^{*}(t)Q(p,t)\right]}{1-\mathcal{L}_{t\to s}\left[\varphi(t)Q(p,t)\right]},
    \label{Qrlt}
\end{eqnarray}
where $Q(p,t)=\int_{0}^{\infty}e^{-pZ_{0}}P(Z_{0}|t)dZ_{0}=\left\langle e^{-pZ_{0}(t)}\right\rangle $ is the characteristic function of the functional when the underlying random walk is free of resetting. 
This equation allows to find the characteristic function of a functional of a random walk under resetting in terms of the characteristic function of the functional of a random walk without resetting and has been recently derived by other means \cite{MeFl05}. Note that in the absence of resetting $\varphi \to 0$ and $\varphi^*\to 1$, so that $\tilde{Q}(p,s)_{r}=\tilde{Q}(p,s)$.

In most cases it is not possible to find $Q(p,t)$ in a closed form; it involves the solution of the Feynman-Kac equation and this is often a difficult task. However, we can deal with the first moments which in many cases is enough to analyze statistical properties like fluctuations and weak ergodic properties of the functionals. Now, we want to relate the moments of the functional in the presence of resetting with the moments of the functional without resetting. To do this, we note that the moment of order $n$ can be found by deriving the characteristic function. The $n$-order moment is given by
\begin{eqnarray}
    \left\langle \tilde{Z}_{r}(s)^{n}\right\rangle =\int_{0}^{\infty}Z_r^{n}\tilde{P}(Z_r,s)dZ_r=(-1)^{n}\frac{\partial^{n}\tilde{Q}(p,s)_{r}}{\partial p^{n}}\bigg|_{p=0}.
    \label{moments}
\end{eqnarray}
Taking the first derivative of Eq. \eqref{Qrlt} with respect to $p$ we find from \eqref{moments} that the first moment is given by
\begin{eqnarray}
    \left\langle \tilde{Z}_r(s)\right\rangle &=&-\frac{\partial \tilde{Q}(p,s)_r}{\partial p^{}}\bigg|_{p=0}=-\frac{\partial}{\partial p^{}}\left(\frac{F^{*}(p,s)}{1-F^{}(p,s)}\right)_{p=0}\\
    &=&-\frac{1}{1-F(0,s)}\frac{\partial F^{*}(p,s)}{\partial p^{}}\bigg|_{p=0}-\frac{F^{*}(0,s)}{\left[1-F(0,s)\right]^{2}}\frac{\partial F^{}(p,s)}{\partial p}\bigg|_{p=0}
\end{eqnarray}
where we have defined $F^*(p,s)=\mathcal{L}_{t\to s}\left[\varphi^{*}(t)Q(p,t)\right]$ and $F(p,s)=\mathcal{L}_{t\to s}\left[\varphi(t)Q(p,t)\right]$. Then, $F(0,\tau)=\varphi(\tau)$ and $F^*(0,\tau)=\varphi^*(\tau)$. Hence,
\begin{eqnarray}
    \left\langle \tilde{Z}_r(s)\right\rangle =\frac{\mathcal{L}_{t\to s}\left[\varphi^{*}(t)\left\langle Z_0(t)\right\rangle \right]}{1-\tilde{\varphi}(s)}+\frac{\mathcal{L}_{t\to s}\left[\varphi^{}(t)\left\langle Z_0(t)\right\rangle \right]}{s\left[1-\tilde{\varphi}(s)\right]}.
    \label{Z1}
\end{eqnarray}
Analogously, taking the second derivative of Eq. \eqref{Qrlt} with respect to $p$, we straightforwardly find
\begin{eqnarray}
 \left\langle \tilde{Z}_r(s)^{2}\right\rangle &=&\frac{2\mathcal{L}_{t\to s}\left[\varphi(t)\left\langle Z_0(t)\right\rangle \right]}{1-\tilde{\varphi}(s)}\left\langle \tilde{Z}_0(s)\right\rangle _{r}\nonumber\\
 &+&\frac{1}{1-\tilde{\varphi}(s)}\left\{ \mathcal{L}_{t\to s}\left[\varphi^{*}(t)\left\langle Z_0(t)^{2}\right\rangle \right]+\frac{1}{s}\mathcal{L}_{t\to s}\left[\varphi(t)\left\langle Z_0(t)^{2}\right\rangle\right]\right\} .
 \label{Z2}
\end{eqnarray}
Following the steps above, we can obtain the corresponding expressions for higher order moments, however to compute the ergodic properties of the observables we need the two first moments only as we show below.

\section{Ergodic properties}
\label{sec:ergodic_properties}
From these two mean values we can derive an expression for the ergodicity breaking parameter which will be useful to study the ergodic properties of $Z(t)$. 
Consider an observable $\mathcal{O}[x(t)]$, a function of the trajectory $\{x(t'); 0\leq t'\leq t\}$. Since $x(t)$ is stochastic in nature, the observable $\mathcal{O}[x(t')]$ will also be fluctuating between the realizations. An observable of the random walk is said to be ergodic if the ensemble average equals the time average $\left\langle \mathcal{O}\right\rangle =\ensuremath{\overline{\mathcal{O}}}$ in the long time limit. This means that if $\mathcal{O}[x(t)]$ is ergodic, then its time average $\ensuremath{\overline{\mathcal{O}}}$ is not a random variable. As a consequence, the limiting PDF of $\ensuremath{\overline{\mathcal{O}}}$, namely $P(\overline{\mathcal{O}},t)$, is a Dirac delta function:
\begin{eqnarray}
P(\overline{\mathcal{O}},t\to\infty)=\delta\left(\overline{\mathcal{O}}-\left\langle \mathcal{\overline{O}}\right\rangle \right).
     \label{lpdf}
 \end{eqnarray} 
 If the observable is integrable with respect to the density $P(x,t|x_0)$, then the ensemble average is given by
\begin{equation}
    \left\langle \mathcal{O}[x(t)]\right\rangle =\int_{-\infty}^{\infty}\mathcal{O}[x]P(x,t|x_0)dx.
\end{equation}
The time average of $\mathcal{O}[x(t)]$ is defined as
\begin{equation}
    \ensuremath{\overline{\mathcal{O}[x(t)]}=}\frac{1}{t}\int_{0}^{t}\mathcal{O}[x(t')]dt'.
    \label{tav}
\end{equation}
For non-ergodic observables, since $\overline{\mathcal{O}}$ is random, its variance $\textrm{Var}(\overline{\mathcal{O}})$ is non-zero in the long time limit. Otherwise, for an ergodic observable $\textrm{Var}(\overline{\mathcal{O}})=0$ in the long time limit. Keeping this in mind, the ergodicity breaking parameter (EB) is defined as
\begin{eqnarray}
\textrm{EB}=\lim_{t\to\infty}\frac{\textrm{Var}(\overline{\mathcal{O}})}{\left\langle \overline{\mathcal{O}}\right\rangle ^{2}}=\lim_{t\to\infty}\frac{\left\langle \overline{\mathcal{O}}^{2}\right\rangle -\left\langle \overline{\mathcal{O}}\right\rangle ^{2}}{\left\langle \overline{\mathcal{O}}\right\rangle ^{2}}.
  \label{EB}
\end{eqnarray}
For ergodic observables, one should have $\textrm{EB}= 0$. 

In the examples below we consider the observable $\mathcal{O}[x(t)]=U[x(t)]$ with $U[x]=|x|$. So that the time average of the observable is from \eqref{tav}
\begin{eqnarray}
    \ensuremath{\overline{\mathcal{O}[x(t)]}=}\frac{1}{t}\int_{0}^{t}U[x(t')]dt'=\frac{Z(t)}{t},
\end{eqnarray}
and so
\begin{eqnarray}
 \ensuremath{\left\langle \overline{\mathcal{O}}\right\rangle =}\frac{\left\langle Z(t)\right\rangle }{t},\quad\ensuremath{\left\langle \overline{\mathcal{O}}^{2}\right\rangle =}\frac{\left\langle Z(t)^{2}\right\rangle }{t^{2}}.   
 \label{o}
\end{eqnarray}
Finally, from \eqref{EB} we can find that the EB parameter of the observable $U[x(t')]$ with $t' \in [0,t]$ is 
\begin{eqnarray}
    \textrm{EB}=\lim _{t\to \infty}\frac{\left\langle Z(t)^{2}\right\rangle }{\left\langle Z(t)\right\rangle ^{2}}-1.
    \label{EB2}
\end{eqnarray}

\section{Absolute area under resetting}
\label{sec:absolute_area}
When one considers the observable $U[x(t')]=|x(t')|$ with $t' \in [0,t]$ then the stochastic functional is known as the absolute area, i.e., the area below the absolute value of the trajectory, 
\begin{eqnarray}
    \mathcal{A}(t)=\int_0^t |x(t')|dt'.
    \label{aad}
\end{eqnarray}
 The EB parameter of the observable $|x(t)|$ is then computed from the two first moments of the absolute area, from \eqref{EB2}
\begin{eqnarray}
    \textrm{EB}=\lim _{t\to \infty}\frac{\left\langle \mathcal{A}(t)^{2}\right\rangle }{\left\langle \mathcal{A}(t)\right\rangle ^{2}}-1.
    \label{EB3}
\end{eqnarray}

In Figure 1 we illustrate in a schematic picture a trajectory under resetting and the area below. 
\begin{figure}[h!]
    \centering
    \includegraphics[width=0.65\linewidth]{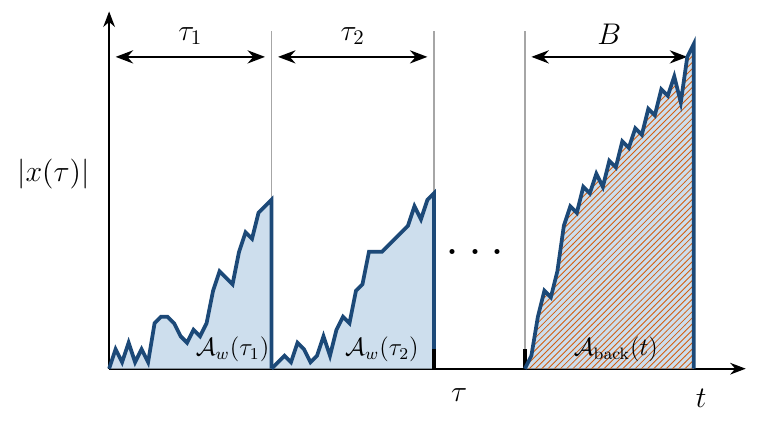}
    \caption{Schematic illustration of a single stochastic trajectory under stochastic resetting. The trajectory is interrupted by resetting events occurring after random waiting times $\tau_1$, $\tau_2$... The blue-shaded region represents the absolute area,  which is the stochastic functional studied throughout this work. The total area is naturally decomposed into the contributions accumulated during each resetting interval together with the contribution from the final backward interval. }
    \label{fig1}
\end{figure}
If the random walk is a SBM the position of the particle satisfies the scaling $x_{\text{SBM}}(t)=x_{\text{BM}}(\eta (t))$ where $\eta(t)=2\int_0^tD(t')dt'$, $D(t)$ is the time-dependent  diffusivity defined by
\begin{eqnarray}
    D(t)=\beta K_\beta t^{\beta -1},\quad  \beta\in(0,2)
    \label{Dsbm}
\end{eqnarray}
 and $K_\beta$ is a positive parameter, and  $x_{\text{BM}}(t)$ is the Wiener process with zero mean and variance $t$. For $0<\beta<1$ the random walk is subdiffusive, for $\beta=1$ we recover the BM with zero mean and variance $2K_1t$, and for $1<\beta<2$ the random walk is superdiffusive. The absolute area below the SBM path up to time $t$ can be expressed as
\begin{eqnarray}
   \mathcal{A}(t)&=&\int_{0}^{t}|x_{\text{SBM}}(\eta(t'))|dt'=\frac{t}{\beta}\int_{0}^{1}z^{\frac{1}{\beta}-1}|x_{\text{BM}}(2K_\beta t^{\beta}z)|dz\nonumber\\
   &=&t^{1+\frac{\beta}{2}}\xi_{\beta},
   \label{sbsbm}
\end{eqnarray}
where we have introduced the definition $z=(t'/t)^{\beta}$ and have made use of the scaling property $x_{\text{BM}}(cz)=\sqrt{c} x_{\text{BM}}(z)$ of the Brownian motion. The random variable $\xi_\beta$ is  defined as 
\begin{eqnarray}
  \xi_\beta=\frac{\sqrt{2K_{\beta}}}{\beta} \int_{0}^{1}z^{\frac{1}{\beta}-1}|x_{\textrm{BM}}(z)|dz.
  \label{xi}
\end{eqnarray}
When $\beta=1$, we have $\mathcal{A}(t)_{\textrm{BM}}=t^{3/2}\xi_{1}$ with $\xi_{1}=\sqrt{2K_1}\int_{0}^{1}|x_{\textrm{BM}}(z)|dz$. 
We now consider the case where the random walk is under the effect of a resetting.

\subsection{Absolute area under Poissonian resetting}
We consider first that the resetting process of the particle's position is governed by a Poisson process. This means that the waiting time PDF between resets is the exponential distribution, this is, $\varphi(t)=re^{-rt}$ where $r$ is the reset rate. 
Inserting the expressions for $\tilde{\varphi} (s)$ and $\tilde{\varphi}^* (s)$ into \eqref{Z1}, the first moment for the absolute area under resetting in the Laplace space is given by
\begin{eqnarray}
    \left\langle \tilde{\mathcal{A}}_r(s)\right\rangle =\left(1+\frac{r}{s}\right)^{2}\left\langle \tilde{\mathcal{A}}_{0}(s+r)\right\rangle
    \label{a1sc}
\end{eqnarray}
where we have made use of the shift property. Using the property of Laplace transform of the convolution product, we can perform the Laplace inversion to the above expression. We obtain 
\begin{eqnarray}
    \left\langle \mathcal{A}_r(t)^{}\right\rangle =\int_{0}^{t}\left\langle \mathcal{A}_{0}(t')^{}\right\rangle e^{-rt'}\lambda(t-t')dt'
\end{eqnarray}
where
$$
\lambda(t)=\delta(t)+2r+r^{2}t
$$
and $\delta(\cdot)$ is the Dirac delta function.
This expression allows to write the first moment of the absolute area under a random walk with exponential resetting in terms of the mean absolute area under a random walk without resetting. 
Likewise, from \eqref{Z2} we write for the second moment
\begin{eqnarray}
    \left\langle \tilde{\mathcal{A}_r}(s)^{2}\right\rangle =2r\left(1+\frac{r}{s}\right)^{3}\left\langle \tilde{\mathcal{A}}_{0}(s+r)\right\rangle^{2}+\left(1+\frac{r}{s}\right)^{2}\left\langle \tilde{\mathcal{A}}_{0}(s+r)^{2}\right\rangle .
    \label{a2sc}
\end{eqnarray}
Performing the Laplace inversion we get
\begin{eqnarray}
    \left\langle \mathcal{A}_r(t)^{2}\right\rangle&=&2r\int_{0}^{t}e^{-rt'}\psi(t-t')dt'\int_{0}^{t'}\left\langle \mathcal{A}_{0}(t'-t'')\right\rangle \left\langle \mathcal{A}_{0}(t'')\right\rangle dt''\\\nonumber
    &+&\int_{0}^{t}\left\langle \mathcal{A}_{0}(t')^{2}\right\rangle e^{-rt'}\lambda(t-t')dt'
\end{eqnarray}
where
$$
\psi (t)=\delta(t)+3r+3r^{2}t+\frac{r^{3}}{2}t^{2}.
$$
Now we find the limiting behavior for the two first moments of the absolute area both in the short and long time limits. To do this we work in the Laplace space. In the short time limit, $t$ is much lower than the characteristic time scale $1/r$ which is equivalent to consider $s\gg r$. On the other hand, in the long time limit $t$ is much higher than  $1/r$, so that $s\ll r$. Considering $s\gg r$ and $s\ll r$ in \eqref{a1sc} and \eqref{a2sc} we find for the first moment under exponential resetting 
\begin{eqnarray}
    \left\langle \mathcal{A}_r(t)\right\rangle \approx\left\{ \begin{array}{cc}
\left\langle \mathcal{A}_{0}(t)\right\rangle , & t\ll1/r\\
\left\langle \mathcal{\tilde{A}}_{0}(r)\right\rangle r^{2}t, & t\gg1/r
\end{array}\right. .
\label{m1lt}
\end{eqnarray}
We see that the first moment increases linearly with time in the long time limit. For the second moment we note that in most cases $\left\langle \mathcal{\tilde{A}}_{0}(s)\right\rangle ^{2}\ll\left\langle \mathcal{\tilde{A}}_{0}(s)^{2}\right\rangle $ when $s$ is large so that the second term of the right hand side in \eqref{a2sc} is much higher than the first term. Hence,
\begin{eqnarray}
    \left\langle \mathcal{A}_r(t)^{2}\right\rangle \approx\left\{ \begin{array}{cc}
\left\langle \mathcal{A}_{0}(t)^{2}\right\rangle , & t\ll1/r\\
\left\langle \mathcal{\tilde{A}}_{0}(r)\right\rangle ^{2}r^{4}t^{2}, & t\gg1/r
\end{array}\right. .
\label{aa2g}
\end{eqnarray}
The two first moments of the absolute area without resetting, $\left\langle \mathcal{A}_{0}(t)\right\rangle $ and $\left\langle \mathcal{A}_{0}(t)^{2}\right\rangle$, can be written in the general form
\begin{eqnarray}
    \left\langle \mathcal{A}_{0}(t)\right\rangle =C_{1}t^{\gamma},\quad\left\langle \mathcal{A}_{0}(t)^{2}\right\rangle =C_{2}t^{2\gamma},
    \label{aceros}
\end{eqnarray}
with $1<1+\frac{\beta}{2}<2$. \cite{MeFlCr26}. For example, for BM with $\left\langle x(t)^{2}\right\rangle _{\text{BM}}=2Dt$ 
one has
\begin{eqnarray}
 C_{1,\textrm{BM}}&=&\frac{4}{3}\sqrt{\frac{D}{\pi}}\nonumber\\
 C_{2,\textrm{BM}}&=&\frac{3D}{4}\nonumber\\
 \gamma_{\textrm{BM}}&=&3/2
 \label{BM}
\end{eqnarray}
while for SBM 
one has
\begin{eqnarray}
C_{1,\textrm{SBM}}&=&\frac{4}{2+\beta}\sqrt{\frac{K}{\pi}}\nonumber\\
C_{2,\textrm{SBM}}&=&\frac{4K}{(\beta+1)(\beta+2)}\left[1+\frac{1}{2\sqrt{\pi}}\frac{\Gamma\left(\frac{1}{2}+\frac{1}{\beta}\right)}{\Gamma\left(2+\frac{1}{\beta}\right)}\right]\nonumber\\
\gamma_{\textrm{SBM}}&=&1+\beta/2
\label{SBM}
\end{eqnarray}
Once $C_1$, $C_2$ and $\gamma$ are known one can compute explicitly the two first moments of the absolute area under Poissonian resetting using \eqref{m1lt} and \eqref{aa2g}.  Note that in the absence of resetting the EB parameter is not zero in general 
$\textrm{EB}_0=\lim_{t\to\infty}\left\langle \mathcal{A}_{0}(t)^{2}\right\rangle /\left\langle \mathcal{A}_{0}(t)\right\rangle ^{2}-1\neq 0$ \cite{MeFlCr26} which means that the observable $|x(t)|$ is not ergodic in the absence of resetting. However, from \eqref{m1lt} and \eqref{aa2g} we find that, regardless of the random walk, the EB parameter for the absolute area swept by a random walk under resetting is from \eqref{EB3}
\begin{eqnarray}
    \textrm{EB}_r=\lim_{t\to\infty}\frac{\left\langle \mathcal{A}_r(t)^{2}\right\rangle} {\left\langle \mathcal{A}_r(t)\right\rangle ^{2}}-1=0
    \label{EBr}
\end{eqnarray}
which means that the observable $|x(\tau)|$ is ergodic when the random walk is under the effect of a Poissonian resetting. In consequence, the Poissonian resetting is a mechanism that induces ergodicity \cite{MeFl05}.

Finally, we can also obtain the general temporal dependence for the first and second moments. 
From \eqref{a1sc} and performing the Laplace transform to \eqref{aceros}
we obtain
$
\left\langle \tilde{\mathcal{A}_r}(s)\right\rangle =C_{1}\Gamma(1+\gamma)/s^{2}(s+r)^{\gamma-1}
$. Using the formula (see Eq. 2.1.2.1 in \cite{pr92})
\begin{eqnarray}
    \mathcal{L}_{s\to t}^{-1}\left[\frac{1}{s^{\mu}(s+r)^{\lambda-1}}\right]=\frac{t^{\mu-2+\lambda}}{\Gamma(\mu-1+\lambda)}M(\lambda-1,\mu-1+\lambda;-rt)
\label{TL}
\end{eqnarray}
for the Laplace inversion, where $M(a,b;z)$ is the Kummer's (confluent hypergeometric) function, we find the exact explicit dependence of the mean absolute area with the observation time
\begin{eqnarray}
    \left\langle \mathcal{A}_r(t)\right\rangle =C_1t^{\gamma}M(\gamma-1,1+\gamma,-rt).
    \label{A2ex}
\end{eqnarray}
Note that if $r=0$ then $M(a,b;0)=1$ and the previous result reduces to \eqref{aceros}.  Note also that for $rt\gg1$ and using Eq. 13.1.5 of Ref. \cite{ab64} we have $M(\gamma-1,1+\gamma,-rt)\approx \Gamma(1+\gamma)(rt)^{1-\gamma} $ so that \eqref{A2ex} coincides with the second line in \eqref{m1lt}.   

\begin{figure}[h!]
    \centering
    \includegraphics[width=0.95\linewidth]{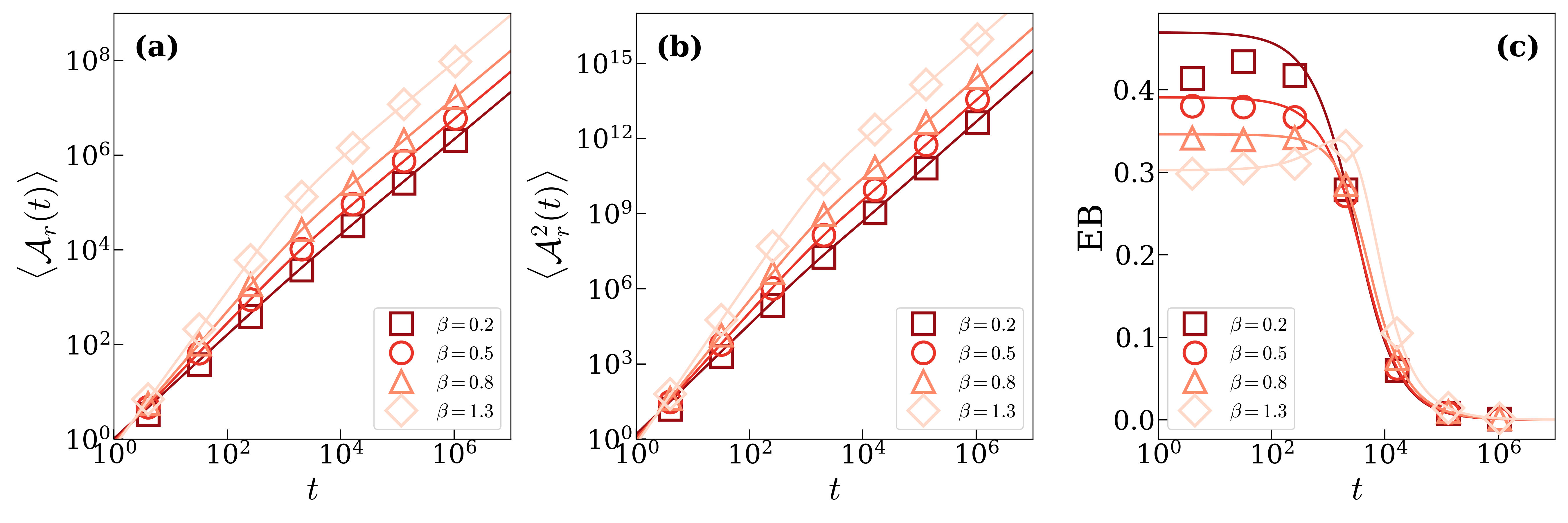}
    \caption{Time evolution of the first moment (a), second moment (b), and EB parameter (c) of the absolute area $\mathcal{A}_r(t)$ under Poissonian resetting. Symbols correspond to stochastic simulations ($N=10^4$ trajectories), while solid lines denote the analytical predictions from Eqs. \eqref{A2ex} and \eqref{A2exx}. Results are shown for SBM with resetting rate $r=10^{-3}$. The long-time decay of the EB parameter towards zero demonstrates that Poissonian resetting restores ergodicity.}
    \label{fig2}
\end{figure}

Analogously, using \eqref{TL} and the Laplace transform to Eqs.  \eqref{aceros}, Eq.  \eqref{a2sc} yields, after Laplace inversion,
\begin{eqnarray}
    \left\langle \mathcal{A}_r(t)^{2}\right\rangle &=&2rC_{1}^{2}\frac{\Gamma(1+\gamma)^{2}}{\Gamma(2+2\gamma)}t^{1+2\gamma}M\left(-1+2\gamma,2+2\gamma;-rt\right)\nonumber\\
    &+&C_{2}t^{2\gamma}M\left(-1+2\gamma,1+2\gamma;-rt\right).
    \label{A2exx}
\end{eqnarray}
It is easy to check that performing the long time limit to the above result yields we recover \eqref{aa2g}. The case for BM was studied in \cite{Ho19}. 
Since EB = 0, from \eqref{lpdf} the PDF of the absolute area under Poissonian resetting converges to  
\begin{eqnarray}
P(\mathcal{A}_r,t)\underset{t\gg1/r}{\approx}\delta\left(\mathcal{A}_r-\left\langle \mathcal{\tilde{A}}_{0}(r)\right\rangle r^{2}t\right)
\end{eqnarray}
in the long time limit. This is in agreement with the general result found in \cite{MeFl05}. In Figure \ref{fig2} we plot the mean values of the absolute area and the squared absolute area  under Poissonian resetting. Theoretical values are found in Eqs. \eqref{A2ex} and\eqref{A2exx}. In panel c) we see how the EB parameter defined in \eqref{EBr}  goes to zero in the long time limit.

\subsection{Absolute area under power-law resetting}
Let us consider that the resetting times are drawn from a fat-tailed PDF $\varphi (t)\sim t^{-1-\alpha}$ with $0<\alpha\leq 2$. It is not difficult to show that for $0<\alpha<1$ all the moments of $\varphi (t)$ are divergent, while if $1<\alpha<2$, only the first moment $\left\langle t_R\right\rangle =\int_0^{\infty}t\varphi(t)dt$ is finite. Using the Tauberian theorem \cite{Feller}, $\varphi (t)$ admits the following expansion in Laplace space for small $s$
\begin{eqnarray}
   \tilde{\varphi}(s)\underset{s\to0}{\approx}\left\{ \begin{array}{ll}
1-b_{\alpha}s^{\alpha}+..., & 0<\alpha<1\\
1-\left\langle t_R\right\rangle s+..., & 1<\alpha<2
\end{array}\right. 
\label{FI}
\end{eqnarray}
where
$b_{\alpha}$ depends on the details of the specific expression for $\varphi (t)$.
In particular, let us consider that the resetting time PDF follows the fat-tailed density
\begin{eqnarray}
    \varphi (t)=\frac{\alpha t_0^\alpha}{t^{1+\alpha}},\quad t>t_0
    \label{pl}
\end{eqnarray}                                                        
and 0 otherwise. In this case $b_{\alpha}=t_{0}^{\alpha}|\Gamma(1-\alpha)|$ and $\left\langle t_{R}\right\rangle =\alpha t_{0}/(\alpha-1)$. Note that the limit $\alpha\to 0$ here corresponds to the absence of resetting because in this limit $\varphi\to 0$ and $\varphi^*\to 1$. Since for \eqref{pl} it is not possible to compute analytically the expressions for the two first moments in real time, we need to explore their behavior in the long time limit. To do this we need to deal with integrals of the form (see Chapter 13 of \cite{ab64})
\begin{eqnarray}
    \int_{t_{0}}^{\infty}e^{-st}t^{\gamma}dt=t_{0}^{\gamma+1}e^{-st_{0}}U\left(1,2+\gamma,st_{0}\right)\nonumber\\
    \int_{0}^{t_{0}}e^{-st}t^{\gamma}dt=\frac{t_{0}^{\gamma+1}}{\gamma+1}e^{-st_{0}}M\left(1,2+\gamma,st_{0}\right)
    \label{K}
\end{eqnarray}
where $U(a,b,z)$ is the Kummer $U$ function. Hence, the terms with Laplace transform in \eqref{Z1} 
follow as
$$
\mathcal{L}_{t\to s}\left[\varphi^{*}(t)\left\langle \mathcal{A}_{0}(t)\right\rangle \right]=C_{1}t_{0}^{\gamma+1}e^{-st_{0}}\left[\frac{1}{\gamma+1}M\left(1,2+\gamma,st_{0}\right)+U\left(1,2+\gamma-\alpha,st_{0}\right)\right]
$$
and
$$
\mathcal{L}_{t\to s}\left[\varphi(t)\left\langle \mathcal{A}_{0}(t)\right\rangle \right]=\alpha C_{1}t_{0}^{\gamma}e^{-st_{0}}U\left(1,1+\gamma-\alpha,st_{0}\right).
$$
To analyze the long time limit $st_0\ll 1$ 
we need to know how the Kummer functions behave when the argument is small. The function $U(1,b,z)$ behaves different depending on the values of $b$. The leading terms in the limit $z\ll 1$ are (see Chapter 13 of \cite{ab64})
\begin{eqnarray}
U(1,b,z)\underset{z\ll1}{\approx}\left\{ \begin{array}{lll}
\frac{1}{1-b}+\frac{z}{b(1-b)}+\dots,\quad & b<0,\\
\frac{1}{1-b}-\frac{\Gamma\left(b\right)}{1-b}z^{1-b}+\dots,\quad & 0<b<1,\\
\frac{\Gamma\left(b-1\right)}{z^{b-1}}+\dots,\quad & b>1
\end{array}\right.
\label{Us0}
\end{eqnarray}
and
$$
M(a,b,z)\underset{z\ll1}{\approx}1+\frac{az}{b}+...
$$
Hence,
\begin{eqnarray}
  \mathcal{L}_{t\to s}\left[\varphi^{*}(t)\left\langle \mathcal{A}_{0}(t)\right\rangle \right]\underset{st_{0}\ll1}{\approx}C_{1}t_{0}^{1+\gamma}\frac{\Gamma(1+\gamma-\alpha)}{(st_{0})^{1+\gamma-\alpha}}  
  \label{la1}
\end{eqnarray}
and
\begin{eqnarray}
    \mathcal{L}_{t\to s}\left[\varphi^{}(t)\left\langle \mathcal{A}_{0}(t)\right\rangle \right]\underset{st_{0}\ll1}{\approx}\alpha C_{1}t_{0}^{\gamma}\left\{ \begin{array}{cc}
\frac{1}{\alpha-\gamma}, & 1+\frac{\beta}{2}<\alpha<2\\
\frac{\Gamma(\gamma-\alpha)}{(st_{0})^{\gamma-\alpha}}, & \alpha<1+\frac{\beta}{2}
\end{array}\right.
\label{la2}
\end{eqnarray}
With the help of \eqref{FI} and with \eqref{la1} and \eqref{la2}, we find
\begin{eqnarray}
    \left\langle \tilde{\mathcal{A}}_r(s)\right\rangle \underset{st_{0}\ll1}{\approx}\left\{ \begin{array}{cc}
\frac{C_{1}t_{0}^{\alpha}\gamma\Gamma(\gamma-\alpha)}{b_{\alpha}s^{1+\gamma}}, & 0<\alpha<1\\
\frac{C_{1}t_{0}^{\alpha}\gamma\Gamma(\gamma-\alpha)}{\left\langle t_{R}\right\rangle s^{2+\gamma-\alpha}}, & 1<\alpha<1+\frac{\beta}{2}\\
\frac{C_{1}t_{0}^{\gamma}\alpha}{\left\langle t_{R}\right\rangle (\alpha-\gamma)s^{2}}, & 1+\frac{\beta}{2}<\alpha<2
\end{array}\right.
\label{As1}
\end{eqnarray}
which after Laplace inversion yields
\begin{eqnarray}
    \left\langle \mathcal{A}_r(t)\right\rangle \underset{t\gg t_{0}}{\approx}\left\{ \begin{array}{cc}
\frac{C_{1}\Gamma(\gamma-\alpha)}{\Gamma(1-\alpha)\Gamma(\gamma)}t^{\gamma}, & 0<\alpha<1\\
\frac{C_{1}t_{0}^{\alpha-1}\gamma(\alpha-1)}{\alpha(1+\gamma-\alpha)(\gamma-\alpha)}t^{1+\gamma-\alpha}, & 1<\alpha<1+\frac{\beta}{2}\\
\frac{C_{1}t_{0}^{\gamma-1}(\alpha-1)}{\alpha-\gamma}t, & 1+\frac{\beta}{2}<\alpha<2
\end{array}\right. .
\label{A1tf}
\end{eqnarray}
We proceed analogously with the second moment, using \eqref{aceros} and \eqref{Z2}. For $1<1+\frac{\beta}{2}<3/2$ we have 
\begin{eqnarray}
    \left\langle \mathcal{A}_r(t)^{2}\right\rangle \underset{t\gg t_{0}}{\approx}\left\{ \begin{array}{cc}
\left[\frac{\alpha C_{1}^{2}\Gamma(\gamma-\alpha)^{2}}{\Gamma(1-\alpha)}+C_{2}\Gamma(2\gamma-\alpha)\right]\frac{t^{2\gamma}}{\Gamma(1-\alpha)\Gamma(2\gamma)}, & 0<\alpha<1\\
\frac{2\gamma C_{2}t_{0}^{\alpha-1}(\alpha-1)}{\alpha(1+2\gamma-\alpha)(2\gamma-\alpha)}t^{1+2\gamma-\alpha}, & 1<\alpha<1+\beta\\
\frac{C_{1}^{2}t_{0}^{2\gamma-2}(\alpha-1)^{2}}{\left(\alpha-\gamma\right)^{2}}t^{2}, & 1+\beta<\alpha<2
\end{array}\right.
\label{A2tf}
\end{eqnarray}
while for $3/2\leq1+\frac{\beta}{2}<2$ we have
\begin{eqnarray}
    \left\langle \mathcal{A}_{r}(t)^{2}\right\rangle \underset{t\gg t_{0}}{\approx}\left\{ \begin{array}{cc}
\left[\frac{\alpha C_{1}^{2}\Gamma(\gamma-\alpha)^{2}}{\Gamma(1-\alpha)}+C_{2}\Gamma(2\gamma-\alpha)\right]\frac{t^{2\gamma}}{\Gamma(1-\alpha)\Gamma(2\gamma)}, & 0<\alpha<1\\
\frac{2\gamma C_{2}t_{0}^{\alpha-1}(\alpha-1)}{\alpha(1+2\gamma-\alpha)(2\gamma-\alpha)}t^{1+2\gamma-\alpha}, & 1<\alpha<2\\
\end{array}\right. .
\label{A2tf2}   
\end{eqnarray}
The EB parameter for the absolute area follows from \eqref{A1tf}, \eqref{A1tf}
 and \eqref{A2tf2}. The resulting expressions are
 \begin{eqnarray}
     \textrm{EB}_{r}\left(0<\beta<1\right)=\left\{ \begin{array}{cc}
\frac{\Gamma\left(1+\frac{\beta}{2}\right)^{2}}{\Gamma\left(2+\beta\right)}\left[\alpha+\frac{C_{2}\Gamma\left(2+\beta-\alpha\right)\Gamma\left(1-\alpha\right)}{C^{2}_{1}\Gamma\left(1+\frac{\beta}{2}-\alpha\right)^{2}}\right]-1, & 0<\alpha<1\\
\frac{2C_{2}\alpha\left(2+\frac{\beta}{2}-\alpha\right)^{2}\left(1+\frac{\beta}{2}-\alpha\right)^{2}}{C^{2}_{1}\left(1+\frac{\beta}{2}\right)(\alpha-1)\left(3+\beta-\alpha\right)(2+\beta-\alpha)}\left(\frac{t}{t_{0}}\right)^{\alpha-1}-1, & 1<\alpha<1+\frac{\beta}{2}\\
\frac{(2+\beta)C_{2}\left(1+\frac{\beta}{2}-\alpha\right)^{2}}{\alpha\left(3+\beta-\alpha\right)\left(2+\beta-\alpha\right)C^{2}_{1}(\alpha-1)}\left(\frac{t}{t_{0}}\right)^{1+\beta-\alpha}-1, & 1+\frac{\beta}{2}<\alpha<1+\beta\\
0, &1+\beta<\alpha<2
\end{array}\right.
\label{EBr1}
 \end{eqnarray}
for a subdiffusive random walk and
\begin{eqnarray}
    \textrm{EB}_{r}\left(1<\beta<2\right)=\left\{ \begin{array}{cc}
\frac{\Gamma\left(1+\frac{\beta}{2}\right)^{2}}{\Gamma\left(2+\beta\right)}\left[\alpha+\frac{C_{2}\Gamma\left(2+\beta-\alpha\right)\Gamma\left(1-\alpha\right)}{C^{2}_{1}\Gamma\left(1+\frac{\beta}{2}-\alpha\right)^{2}}\right]-1, & 0<\alpha<1\\
\frac{2C_{2}\alpha\left(2+\frac{\beta}{2}-\alpha\right)^{2}\left(1+\frac{\beta}{2}-\alpha\right)^{2}}{C^{2}_{1}\left(1+\frac{\beta}{2}\right)(\alpha-1)\left(3+\beta-\alpha\right)(2+\beta-\alpha)}\left(\frac{t}{t_{0}}\right)^{\alpha-1}-1, & 1<\alpha<1+\frac{\beta}{2} \\
\frac{(2+\beta)C_{2}\left(1+\frac{\beta}{2}-\alpha\right)^{2}}{\alpha\left(3+\beta-\alpha\right)\left(2+\beta-\alpha\right)C^{2}_{1}(\alpha-1)}\left(\frac{t}{t_{0}}\right)^{1+\beta-\alpha}-1, & 1+\frac{\beta}{2}<\alpha<2
\end{array}\right.
\label{EBr2}
\end{eqnarray}
for a superdiffusive random walk. For  the specific case of BM ($\beta=1$) using the parameters given in \eqref{BM} we find from \eqref{EB3}, \eqref{A1tf},  \eqref{A2tf} and \eqref{A2tf2}  that for 
 $0<\alpha<1$
 \begin{eqnarray}
     \textrm{EB}_{r}(0<\alpha<1)=\frac{\pi}{8}\left[\alpha+\pi\left(\frac{3}{4}\right)^{3}\frac{\Gamma(3-\alpha)\Gamma(1-\alpha)}{\Gamma(3/2-\alpha)^{2}}\right]-1
 \end{eqnarray}
 while for $1<\alpha<2$ the parameter EB$_r$ is time-dependent:
 \begin{eqnarray}
     \textrm{EB}_{r}(1<\alpha<2)\underset{t\gg t_0}{\approx}\left\{ \begin{array}{cc}
A(t/t_{0})^{\alpha-1}, & 1<\alpha<3/2\\
B(t/t_{0})^{2-\alpha}, & 3/2<\alpha<2
\end{array}\right.
 \end{eqnarray}
where
$$
A=\pi\left(\frac{3}{4}\right)^{2}\frac{\left(3/2-\alpha\right)^{2}\left(5/2-\alpha\right)^{2}}{(4-\alpha)(3-\alpha)(\alpha-1)}
$$
and
$$
B=3\pi\left(\frac{3}{4}\right)^{3}\frac{\left(3/2-\alpha\right)^{2}}{\alpha(4-\alpha)(3-\alpha)(\alpha-1)}.
$$
It is interesting to note the variety of temporal scalings of the first two moments and the EB as a function of the exponent values $\alpha$ and $\beta$. We see up to three different regimes. For $0<\alpha<1$ (the waiting times PDF has all moments divergent) EB is constant, which means that the resetting process is governed by very few extreme events. For $1<\alpha<2$ (the waiting times PDF has only the first moment finite) EB is time-dependent, which means that there are many resets of large duration (large area). Within the regime $1<\alpha<2$ there is a singular situation if the movement is subdiffusive ($0<\beta<1$): EB = 0 if $\alpha>\alpha_c$ where the critical value is $\alpha_c=1+\beta$. This region corresponds to an ergodic phase in which there are many resets of typical duration (i.e., typical value for the area). The slow motion of a subdiffusive random walk is a requirement for sustaining the ergodicity for the absolute area across the  range $1+\beta<\alpha<2$. 

\begin{figure}[h!]
    \centering
    \includegraphics[width=0.99\linewidth]{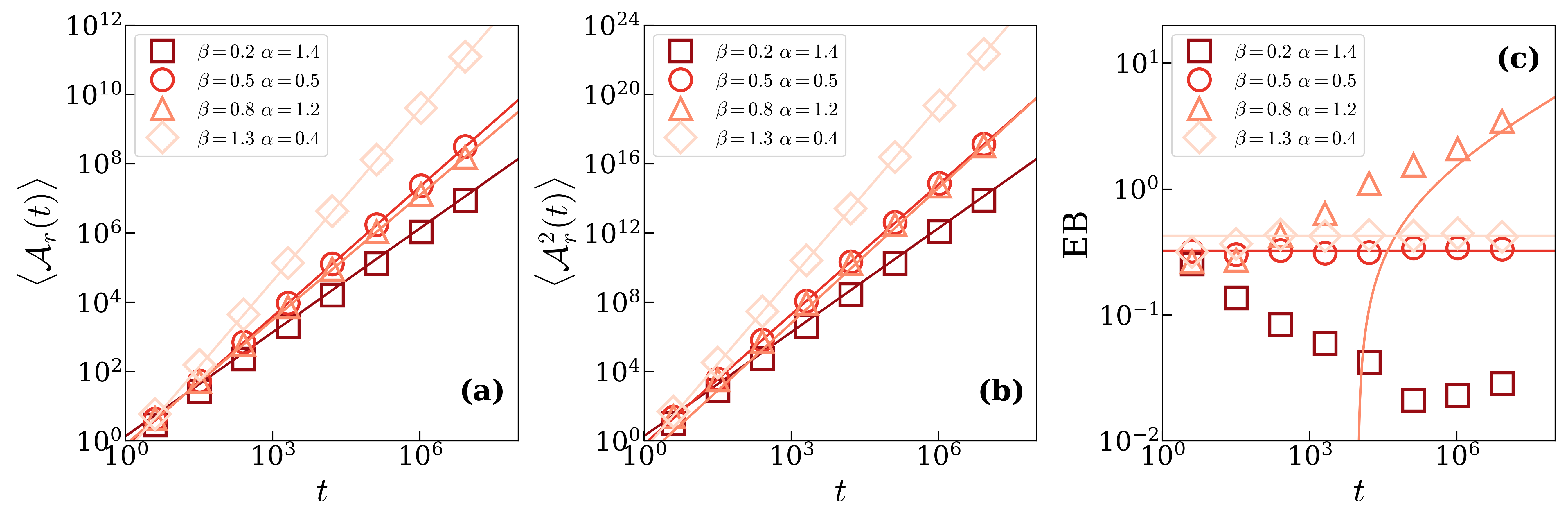}
    \caption{Time evolution of the first moment (a), second moment (b), and EB parameter (c) of the absolute area $\mathcal{A}_r(t)$ under power-law resetting. Symbols represent simulations and solid lines the analytical asymptotic predictions from Eqs. \eqref{A1tf}, \eqref{A2tf}, \eqref{A2tf2}, \eqref{EBr1} or \eqref{EBr2}. For the $\beta=0.2$ and $\alpha=1.4$ case, the theoretical prediction is that the EB parameter goes to $0$ in the long time limit.}
    \label{fig3}
\end{figure}

In Figure \ref{fig3} we plot the mean values of the absolute area and the squared absolute area under power law  resetting. Theoretical are found in  Eqs. \eqref{A1tf}, \eqref{A2tf} or \eqref{A2tf2}. In panel c) we see how the EB parameter given in Eqs. \eqref{EBr1} or \eqref{EBr2} evolve differently with time $t$. For example,  triangles represent the case where EB increases with $t$ in the long time limit. This indicates persistent trajectory-to-trajectory fluctuations despite the existence of a finite mean resetting time. In this case resetting is not able to erase fluctuations. Circles correspond to the case where EB is constant in the long time limit. Squares correspond to the case where EB = 0. This is in agreement with the results of Eq. \eqref{EBr1}.

\section{Backward time contribution}
\label{sec:backward} 
To understand the different temporal scalings of the two first moments of the absolute area under power law resetting we analyze here how is the scaling of the absolute area swept during the backward time. 
Since the resetting is a renewal process, after the last reset the particle starts a renewed path of duration $B$ up to the observation time $t$. Then, the absolute area along the backward time is
$$
\mathcal{A}_{\text{back}}(B)=\int_{0}^{B}|x(t')|dt'.
$$
For SBM the above integral can be written as 
\begin{eqnarray}
  \mathcal{A}_{\text{back}}(B)=B^{1+\frac{\beta}{2}}\xi_{\beta},  
  \label{ssbm}
\end{eqnarray}
where $\xi_{\beta}$ is defined in Eq. \eqref{xi}.  Since, $B$ and $\xi_\beta$ are independent random variables, averaging over the ensemble of trajectories one finds 
\begin{eqnarray}
\left\langle \mathcal{A}_{\text{back}}(t)\right\rangle =\left\langle B^{1+\frac{\beta}{2}}\right\rangle \left\langle \xi_{\beta}\right\rangle.  
\label{masbm}
\end{eqnarray}
On the one hand, the fractional moment of the backward time is
$$\left\langle B^{1+\frac{\beta}{2}}\right\rangle =\int_{0}^{t}B^{1+\frac{\beta}{2}}\Psi(B,t)dB.$$
For Poissonian resetting \cite{GoLu01} the PDF of the backward time is given by
\begin{eqnarray}
 \Psi(B,t)=e^{-rB}\delta(t-B)+re^{-rB}\theta(t-B), 
 \label{pdfbe}
\end{eqnarray}
where $\theta (\cdot)$ is the Heaviside step function. Hence, using \eqref{K}, $\left\langle B^{1+\frac{\beta}{2}}\right\rangle =t^{1+\frac{\beta}{2}}h_1(z)$ where $h_1(z)=e^{-z}\left[1+\frac{2z}{4+\beta}M(1,3+\frac{\beta}{2},z)\right]$ with $z\equiv rt$. To find $\left\langle \xi_{\beta}\right\rangle$ we note from \eqref{sbsbm} that the mean absolute area without resetting is given in Eq. \eqref{aceros}, so that 
\begin{eqnarray}
   \left\langle \xi_{\beta}\right\rangle =\frac{4}{2+\beta}\sqrt{\frac{K}{\pi}}.
   \label{xi1}
\end{eqnarray}

Finally, from \eqref{ssbm}, the contribution of the backward time to the mean absolute area is given by
\begin{eqnarray}
    \left\langle \mathcal{A}_{\text{back}}(t)\right\rangle =\frac{4}{2+\beta}\sqrt{\frac{K}{\pi}}t^{1+\frac{\beta}{2}}h_1(z).
\end{eqnarray}
In the long time limit, $rt\gg 1$, using $M(1,b,x)\approx \Gamma(b)x^{1-b}e^x+O(x^{-1})$ for large $x$ we find that $\left\langle \mathcal{A}_{\text{back}}(t)\right\rangle $ scales as $t^0$, in particular
\begin{eqnarray}
\left\langle \mathcal{A}_{\text{back}}(t)\right\rangle \underset{t\to\infty}{\approx}2\sqrt{\frac{K}{\pi}}\frac{\Gamma\left(1+\frac{\beta}{2}\right)}{r^{1+\frac{\beta}{2}}}.
\label{Ab1}
\end{eqnarray}

\begin{figure}[h!]
    \centering
    \includegraphics[width=0.75\linewidth]{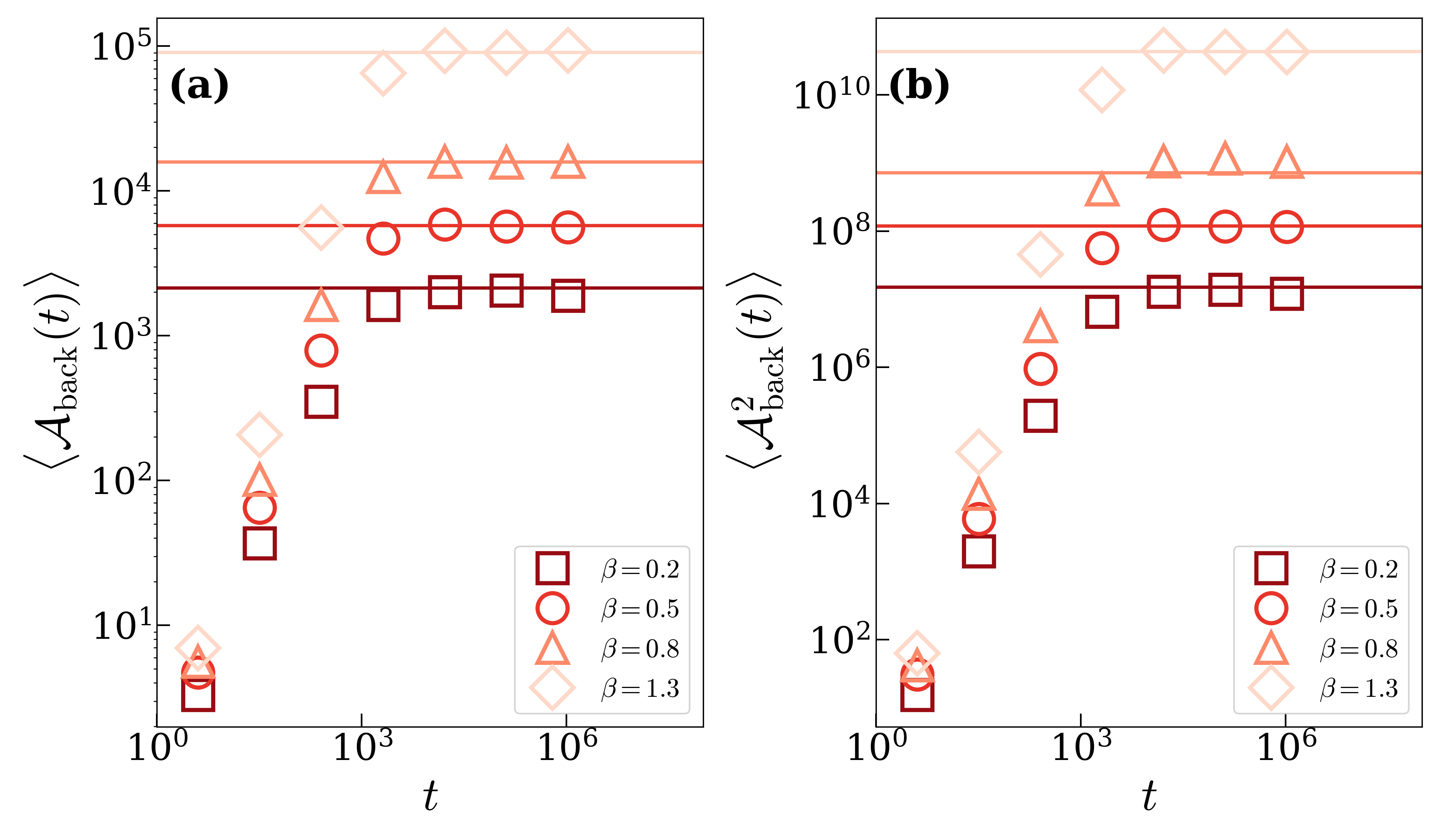}
    \caption{First (a) and second (b) moments of $\mathcal{A}_{\text{back}}(t)$ under Poissonian resetting with $r=10^{-3}$. Symbols denote simulations and solid lines the asymptotic theoretical predictions given by Eqs. \eqref{Ab1} and \eqref{Abr2}. Both moments saturate to a constant in the long time limit.}
    \label{fig4}
\end{figure}
In figure \ref{fig4} we plot the temporal evolution of the two first moments of the backward time contribution to the absolute area for a SBM under Poissonian resetting. The theoretical results are given by Eqs. \eqref{Ab1}
 and \eqref{Abr2}. In both cases the values saturate to a constant value in the long time limit which is in agreement with theoretical results.

For power-law resetting the contribution of the backward time is quite different. For distributions with finite first moment, equilibrium is attained at long times, for the backward time and \cite{GoLu01}
\begin{eqnarray}
\Psi(B,t)\underset{t\to\infty}{\approx}\frac{\varphi^{*}(B)}{\left\langle t_{R}\right\rangle }=\left\{ \begin{array}{cc}
\frac{1}{\left\langle t_{R}\right\rangle }, & B<t_{0}\\
\frac{t^{\alpha}_{0}}{\left\langle t_{R}\right\rangle B^{\alpha}}, & B>t_{0}
\end{array}\right.,\quad1<\alpha<2.
    \label{pdfb1}
\end{eqnarray}

\begin{figure}[h!]
    \centering
    \includegraphics[width=0.75\linewidth]{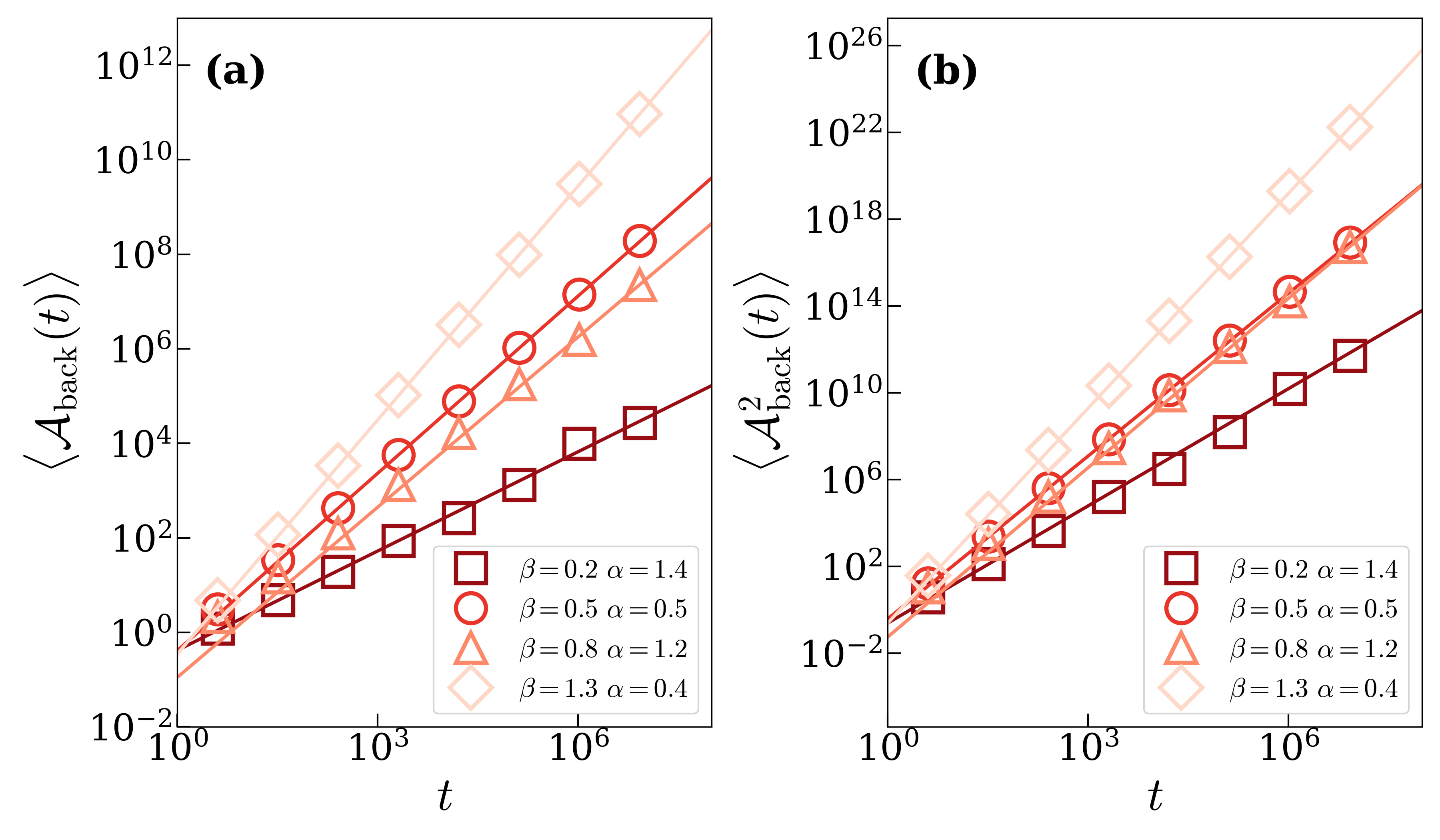}
    \caption{First (a) and second (b) moments of $\mathcal{A}_{\text{back}}(t)$ for different values of $\alpha$ and $\beta$. Symbols denote simulations and solid lines the asymptotic theoretical predictions given by Eqs. \eqref{Absr}, \eqref{aab}, \eqref{abr2} and \eqref{ab2}.}
    \label{fig5}
\end{figure}

 Considering \eqref{pl} and \eqref{masbm} we find that the contribution of the backward time to the mean absolute area for power-law resetting times with finite first moment ($1<\alpha<2$) is given by
\begin{eqnarray}
    \left\langle \mathcal{A}_{\text{back}}(t)\right\rangle \underset{t\to\infty}{\approx}\frac{4}{2+\beta}\sqrt{\frac{K}{\pi}}\frac{\alpha-1}{\alpha t_{0}^{1-\alpha}}\frac{t^{2+\frac{\beta}{2}-\alpha}}{2+\frac{\beta}{2}-\alpha}\sim t^{2+\frac{\beta}{2}-\alpha}.
    \label{Absr}
\end{eqnarray}
Although the PDF of reset times has finite moments, as the Poissonian  resetting, we note that this scaling differs from that obtained for Poissonian resetting. For power-law distribution with diverging moments ($0<\alpha<1$) the PDF of the backward time in the long time limit is given by the generalized arcsine distribution \cite{GoLu01}
\begin{eqnarray}
   \Psi(B,t)\underset{t\to\infty}{\approx} \frac{\sin(\pi\alpha)}{\pi}B^{-\alpha}(t-B)^{-1+\alpha},\quad 0<\alpha<1
   \label{arcsine}
\end{eqnarray}
 and from \eqref{pl} one gets
\begin{eqnarray}
    \left\langle \mathcal{A}_{\text{back}}(t)\right\rangle \underset{t\to\infty}{\approx}\frac{4}{2+\beta}\sqrt{\frac{K}{\pi}}\frac{\Gamma\left(2+\frac{\beta}{2}-\alpha\right)}{\Gamma\left(2+\frac{\beta}{2}\right)\Gamma\left(1-\alpha\right)}t^{1+\frac{\beta}{2}}\sim t^{1+\frac{\beta}{2}}.
    \label{aab}    
\end{eqnarray}
For the second moment we proceed in the same way as we did for the first moment. From \eqref{ssbm} we have
\begin{eqnarray}
    \left\langle \mathcal{A}_{\text{back}}^2\right\rangle =\left\langle B^{2+\beta}\right\rangle \left\langle \xi_{\beta}^2\right\rangle.  
    \label{msmsbm}
\end{eqnarray}
The quantity $\left\langle \xi_{\beta}^2\right\rangle$ is given in Eq. \eqref{SBM} according to \eqref{aceros}, i.e.,
\begin{eqnarray}
    \left\langle \xi_{\beta}^2\right\rangle=\frac{4K}{(\beta+1)(\beta+2)}\left[1+\frac{1}{2\sqrt{\pi}}\frac{\Gamma\left(\frac{1}{2}+\frac{1}{\beta}\right)}{\Gamma\left(2+\frac{1}{\beta}\right)}\right].
    \label{xi2}
\end{eqnarray}
For Poissonian resetting $\left\langle B^{2+\beta}\right\rangle =t^{2+\beta}h_2(z)$,
with $h_{2}(z)=e^{-z}\left[1+\frac{z}{3+\beta}M(1,4+\beta,z)\right]$. Finally, from \eqref{msmsbm} we get
\begin{eqnarray}
\left\langle \mathcal{A}_{\text{back}}(t)^{2}\right\rangle =C_{2,\text{SBM}}h_{2}(z)t^{2+\beta}\underset{t\to\infty}{\approx}C_{2,\text{SBM}}\frac{\Gamma(3+\beta)}{r^{2+\beta}}.
\label{Abr2}
\end{eqnarray}
For power-law resetting with finite mean reset time ($1<\alpha<2$) one has $\left\langle B^{2+\beta}\right\rangle \approx t_{0}^{\alpha}t^{3+\beta-\alpha}/\left\langle t_{R}\right\rangle(3+\beta-\alpha)$ and from \eqref{msmsbm} one finally finds
\begin{eqnarray}
   \left\langle \mathcal{A}_{\text{back}}(t)^{2}\right\rangle \underset{t\to\infty}{\approx}\frac{(\alpha-1)C_{2,\text{SBM}}t_{0}^{\alpha-1}}{\left(3+\beta-\alpha\right)\alpha}t^{3+\beta-\alpha}\sim t^{3+\beta-\alpha}.
    \label{abr2}
\end{eqnarray}
Finally, for power-law resetting with diverging mean reset time ($0<\alpha<1$) the fractional moment of the backward time is $\left\langle B^{2+\beta}\right\rangle \approx \Gamma(3+\beta-\alpha)t^{2+\beta}/\Gamma\left(1-\alpha\right)\Gamma(3+\beta)$ so that from \eqref{msmsbm} we find 
\begin{eqnarray}
    \left\langle \mathcal{A}_{\text{back}}(t)^{2}\right\rangle \underset{t\to\infty}{\approx}\frac{C_{2,\text{SBM}}\Gamma\left(3+\beta-\alpha\right)}{\Gamma\left(3+\beta\right)\Gamma\left(1-\alpha\right)}t^{2+\beta}\sim t^{2+\beta}.
    \label{ab2}
\end{eqnarray}
In figure \ref{fig5} we plot the temporal evolution of the two first moments of the backward time contribution to the absolute area for a SBM under power-law resetting.
Unlike Poissonian resetting, both moments increase with time as a power-law.

We are interested in finding the PDF for $\mathcal{A}_{\text{back}}$. Since $B$ and $\xi_\beta$ are independent random variables, from \eqref{ssbm} we can write the PDF of $\mathcal{A}_{\text{back}}$ as 
\begin{eqnarray}
    P_{\mathcal{A}_{\text{back}}}(a,t)=\int^{\infty}_{0}P(\xi_{\beta})d\xi_{\beta}\int^{t}_{0}\Psi(B,t)\delta\left(a-B^{1+\frac{\beta}{2}}\xi_{\beta}\right)dB.
    \label{pdfamaxe}
\end{eqnarray}
For Poissonian resetting, inserting Eq. \eqref{pdfbe} into Eq. \eqref{pdfamaxe}, we find that the PDF of $\mathcal{A}_{\text{back}}$ reaches a stationary state in the long time limit, this is
\begin{eqnarray}
    P_{\mathcal{A}_{\text{back}}}(a,\infty)=r\int^{\infty}_{0}\frac{e^{-rB}}{B^{1+\frac{\beta}{2}}}P\left(\frac{a}{B^{1+\frac{\beta}{2}}}\right)dB.
\end{eqnarray}
To compute explicitly $P_{\mathcal{A}_{\text{back}}}(a,\infty)$ we need to know the PDF of $\xi_\beta$, $P(\cdot)$. 

\begin{figure}[h!]
    \centering
    \includegraphics[width=1\linewidth]{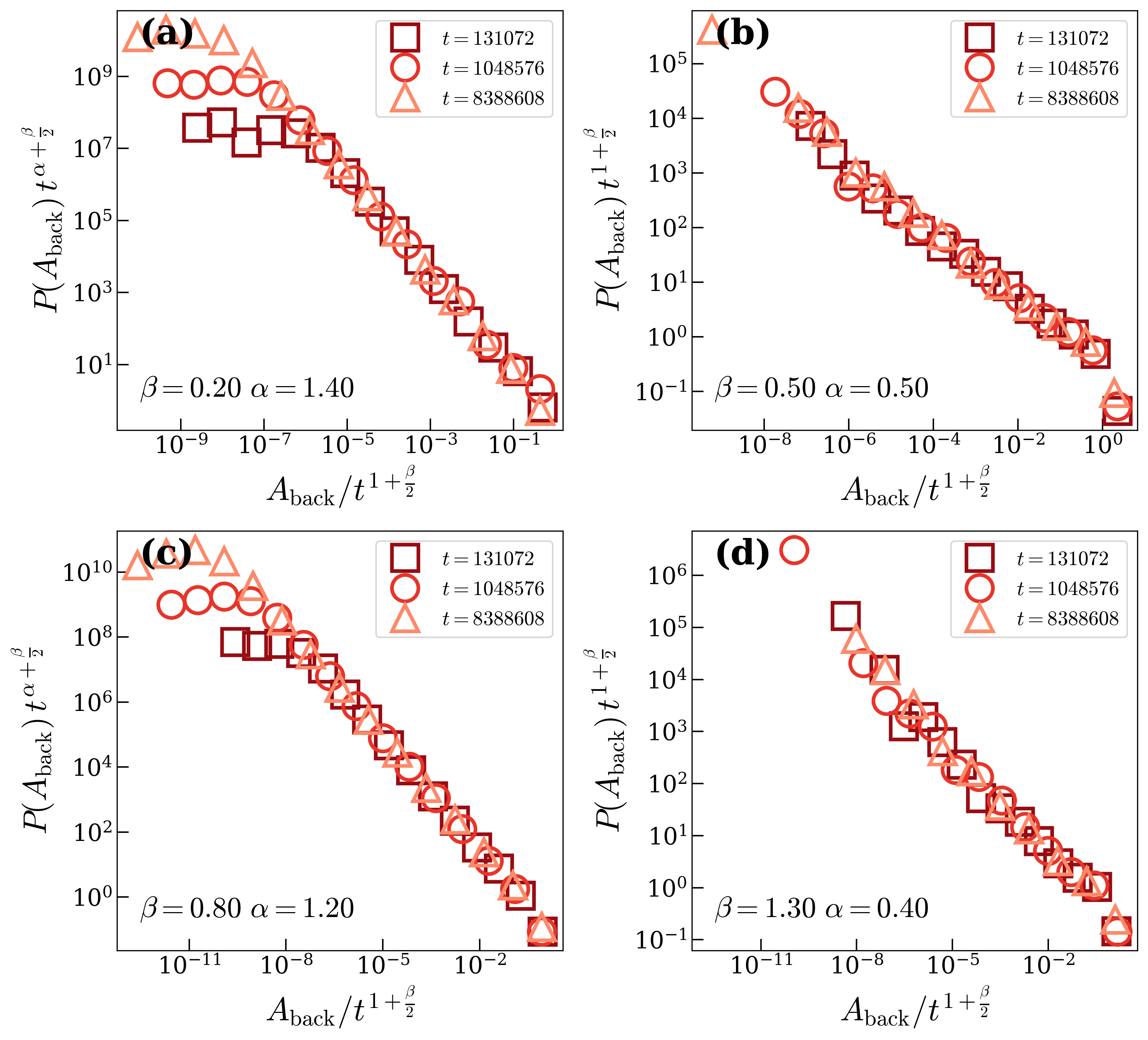}
    \caption{Scaling collapse of the probability density of the backward time contribution to the absolute area $\mathcal{A}_{\text{back}}(t)$ for a SBM under power-law resetting. The rescaled distributions obtained at different observation times collapse onto the predicted scaling function, confirming the asymptotic scaling forms derived in Eqs. \eqref{sf1} and \eqref{sf2}.}
    \label{fig6}
\end{figure}

For power law resetting with $0<\alpha<1$, $\mbox{PDF}(B,t)$ is given by \eqref{arcsine} and then $P_{\mathcal{A}_{\text{back}}}(a,t)$ can be written in the scaling form
\begin{eqnarray}
t^{1+\frac{\beta}{2}}P_{\mathcal{A}_{\text{back}}}(a,t)\underset{t\to\infty}{\approx}h_{\alpha,\beta}\left(\frac{a}{t^{1+\frac{\beta}{2}}}\right)
\label{sf1}
\end{eqnarray}
where the scaling function $h_{\alpha,\beta}(\cdot)$ depends explicitly on the PDF of $\xi_\beta$ as
$$
h_{\alpha,\beta}(\rho)=\frac{\sin(\pi\alpha)}{\pi\rho^{\alpha+\frac{\beta}{2}}}\int^{\infty}_{\rho}\frac{P(z)z^{\beta/2}}{(z-\rho)^{1-\alpha}}dz,\quad\rho=\frac{a}{t^{1+\frac{\beta}{2}}}.
$$
From this expression one can verify the scalings \eqref{aab} and \eqref{ab2}. Analogously, for $1<\alpha<2$ the PDF of $B$ is given by \eqref{pdfb1} and from \eqref{pdfamaxe} 
we finally find the scaling
\begin{eqnarray}
t^{\alpha+\frac{\beta}{2}}P_{\mathcal{A}_{\text{back}}}(a,t)\underset{\underset{a\gg t^{1+\beta/2}_{0}}{t\gg t_{0}}}{\approx}l_{\alpha,\beta}\left(\frac{a}{t^{1+\frac{\beta}{2}}}\right)
\label{sf2}
\end{eqnarray}
where 
$$
l_{\alpha,\beta}\left(\rho\right)=\frac{2t^{\alpha}_{0}}{(2+\beta)\left\langle t_{R}\right\rangle \rho^{\frac{2\alpha+\beta}{2+\beta}}}\int^{\infty}_{\rho}\frac{P(u)du}{u^{\frac{2(1-\alpha)}{2+\beta}}}.
$$
In figure \ref{fig6} we check the validity of the scaling forms \eqref{sf1} and \eqref{sf2} by plotting numerical simulations results for three different values of the observation time. As can be seen the collapse is visually evident. At $\alpha =1$ (the value at which the backward time PDF changes from the power-law to the arcsine distribution), the PDF of $\mathcal{A}_{\text{back}}(t)$ also changes its scaling form.

\section{Waiting time contribution}
\label{sec:waiting_time}
Here we want to calculate the contribution to the total area of the areas under the segments of the trajectory that take place  during the waiting periods between resetting events (see Figure 1). This corresponds to the first term of the right hand side of Eq. \eqref{contrib} for $Z=\mathcal{A}$. The absolute area swept by a trajectory affected by $N$ resets can be expressed as 
\begin{eqnarray}
  \mathcal{A}_N(t)=  \sum_{k=1}^{N}\mathcal{A}_{w}(\tau_{k}),
  \label{aa}
\end{eqnarray}
where each $\mathcal{A}_{w}(\tau_{k})=\int_0^{\tau_k}|x(t')|dt'$ is the contribution from the absolute area over each waiting period. 

First, we find the characteristic function of $ \mathcal{A}_N(t)$. Let $P_{ \mathcal{A}_N} (a,N,t)$ be the joint probability of $\mathcal{A}_N$ and $N$ in the observation time $t$. Then, $P_{ \mathcal{A}_N} (a,N,t)=P_{ \mathcal{A}_N} (a|N)P(N|t)$, where $P_{ \mathcal{A}_N} (a|N)$ is the probability that $\mathcal{A}_N$ takes the value $a$ given that have occurred $N$ resets and $P(N|t)$ is the probability that there are $N$ resets in the observation time $t$. Summing $P_{ \mathcal{A}_N} (a,N,t)$ over all resets we obtain the marginal probability
\begin{eqnarray}
    P_{\mathcal{A}_{N}}(a,t)=\sum_{N=0}^{\infty}P_{\mathcal{A}_{N}}(a,N,t)=\sum_{N=0}^{\infty}P_{\mathcal{A}_{N}}(a|N)P(N|t),
    \label{renewal}
\end{eqnarray}
which is the PDF of $\mathcal{A}_{N}(t)$. Since $\mathcal{A}_{N}(t)$ is sum of the areas $\mathcal{A}_{w}$ (which are IID random variables) as shown in Eq. \eqref{aa}, the probability of $\mathcal{A}_{N}$ conditioned to $N$ resets is the $N$-time convolution product $P_{\mathcal{A}_{N}}(a|N)=P_{\mathcal{A}_{w}}(a)\ast\overset{N}{\cdots}\ast P_{\mathcal{A}_{w}}(a)$. Introducing the Laplace transform ($a\to p$) we define the characteristic function of  $\mathcal{A}_{N}(t)$ as $Q_{\ensuremath{\mathcal{A}_{N}}}(p,t)=\mathcal{L}_{a\to p}\left[P_{\ensuremath{\mathcal{A}_{N}}}(a,t)\right]=\int_{0}^{\infty}e^{-pa}P_{\ensuremath{\mathcal{A}_{N}}}(a,t)da$. Analogously, we define the characteristic function of $\mathcal{A}_{w}$ as $Q_{\mathcal{A}_{w}}(p)=\int_{0}^{\infty}e^{-pa}P_{\ensuremath{\mathcal{A}_{w}}}(a)da$.  Then, performing the Laplace transform to Eq. \eqref{renewal} we find a relationship between the characteristic functions of $\mathcal{A}_{N}(t)$ and $\mathcal{A}_{w}$
\begin{eqnarray}
    Q_{\mathcal{A}_{N}}(p,t)=\sum_{N=0}^{\infty}Q_{\mathcal{A}_{w}}(p)^{N}P(N|t).
\end{eqnarray}
We can compute now the two first moments of $\mathcal{A}_{N}(t)$ from this equation. Taking the first derivative with respect to $p$ and setting $p=0$ we get
\begin{eqnarray}
    \left\langle \mathcal{A}_{N}(t)\right\rangle &=&-\left(\frac{\partial Q_{\mathcal{A}_{N}}(p,t)}{\partial p}\right)_{p=0}=-\left(\frac{\partial Q_{\mathcal{A}_{w}}(p)}{\partial p}\right)_{p=0}\sum_{N=0}^{\infty}NQ_{\mathcal{A}_{w}}(0)^{N-1}P(N|t)\nonumber\\
    &=&\left\langle \mathcal{A}_{w}\right\rangle \left\langle N(t)\right\rangle 
    \label{man1}
\end{eqnarray}
Analogously, taking the second derivative with respect to $p$ and setting $p=0$ we obtain
\begin{eqnarray}
    \left\langle \mathcal{A}_{N}(t)^{2}\right\rangle &=&\left(\frac{\partial^{2}Q_{\mathcal{A}_{N}}(p,t)}{\partial p^{2}}\right)_{p=0}=\left(\frac{\partial Q_{\mathcal{A}_{w}}(p)}{\partial p}\right)_{p=0}^{2}\sum_{N=0}^{\infty}N(N-1)Q_{\mathcal{A}_{w}}(0)^{N-2}P(N|t)\nonumber\\
    &+&\left(\frac{\partial^{2}Q_{\mathcal{A}_{w}}(p)}{\partial p^{2}}\right)_{p=0}\sum_{N=0}^{\infty}NQ_{\mathcal{A}_{w}}(0)^{N-1}P(N|t)\nonumber\\
    &=&\left\langle \mathcal{A}_{w}\right\rangle ^{2}\left(\left\langle N(t)^{2}\right\rangle -\left\langle N(t)\right\rangle \right)+\left\langle \mathcal{A}_{w}^{2}\right\rangle \left\langle N(t)\right\rangle 
    \label{man2}
\end{eqnarray}
where we have taken into account that $Q_{\mathcal{A}_{w}}(0)=1$. Eqs. \eqref{man1} and \eqref{man2} are relationships between the two first moments of $\mathcal{A}_{N}(t)$ and the two first moments of $\mathcal{A}_{w}(t)$. 

 Consider that the particle follows a SBM. Then, along a waiting period of duration $\tau$ the absolute area without resetting obeys the relation \eqref{sbsbm}, $\mathcal{A}_{w}(\tau)=\tau ^{1+\frac{\beta}{2}}\xi_\beta$, i.e., it is the product of two independent random variables. $\tau$ is the time between consecutive resets and is distributed according to $\varphi (\tau)$ and $\xi_\beta$ is defined in Eq. \eqref{xi}. Then, the PDF of $\mathcal{A}_{w}$ can be obtained by a simple change of variables and is given by
\begin{eqnarray}
    P_{\mathcal{A}_{w}}(a)=\int_{0}^{\infty}d\xi_{\beta}P(\xi_{\beta})\int_{0}^{\infty}d\tau\varphi(\tau)\delta\left(a-\tau^{1+\frac{\beta}{2}}\xi_{\beta}\right)
    \label{Pa0}
\end{eqnarray}
where $P(\xi_\beta)$ is the PDF of $\xi_\beta$. Using \eqref{pl} we find
\begin{eqnarray}
 P_{\mathcal{A}_{w}}(a)=\frac{2\alpha t_{0}^{\alpha}}{2+\beta}a^{-1-\frac{2\alpha}{2+\beta}}\int_{0}^{a/t_{0}^{1+\frac{\beta}{2}}}P(\xi_{\beta})\xi_{\beta}^{\frac{2\alpha}{2+\beta}}d\xi_{\beta}.
 \label{pdfa0}
\end{eqnarray}
It is not difficult to realize that the tail of this PDF is also a power-law. In the long $a$ limit, this is for $a\gg t_0^{1+\beta/2}$, we have 
\begin{eqnarray}
     P_{\mathcal{A}_{w}}(a)\underset{a\gg t_{0}^{1+\beta/2}}{\approx}\frac{2\alpha t_{0}^{\alpha}}{2+\beta}\frac{\left\langle \xi^{\frac{2\alpha}{2+\beta}}\right\rangle }{a^{1+\frac{2\alpha}{2+\beta}}},
     \label{pawap}
\end{eqnarray}
i.e., the tail of the PDF of  $\mathcal{A}_{w}$ behaves as  $P_{\mathcal{A}_{w}}(a)\sim a^{-1-\frac{2\alpha}{2+\beta}}$. Depending on the values of $\alpha$ and $\beta$  the two first moments of this PDF may be finite or divergent, and this defines the different regimes that appear in the expressions given in \eqref{A1tf}, \eqref{A2tf} and \eqref{A2tf2}.  To find the moments let us to introduce the characteristic function $Q_{\mathcal{A}_{w}}(p)$ of $P_{\mathcal{A}_{w}}(a)$. From \eqref{pdfa0},
$$Q_{\mathcal{A}_{w}}(p)=\int _0^\infty e^{-pa}P_{\mathcal{A}_{w}}(a)da=\int_{0}^{\infty}d\xi_{\beta}P(\xi_{\beta})\int_{0}^{\infty}d\tau\varphi(\tau)e^{-p\tau^{1+\frac{\beta}{2}}\xi_{\beta}} 
$$
and the first moment is
\begin{eqnarray}
   \left\langle \mathcal{A}_{w}\right\rangle =-\left(\frac{\partial Q_{\mathcal{A}_{w}}(p)}{\partial p}\right)_{p=0}=\left\langle \xi_{\beta}\right\rangle \int_{0}^{\infty}d\tau\varphi(\tau)\tau^{1+\frac{\beta}{2}}. 
   \label{awm1}
\end{eqnarray}

For the power-law waiting time PDF given in \eqref{pl}, this integral is divergent if $\alpha <1+\beta/2$ and convergent otherwise. Analogously, the second moment reads
\begin{eqnarray}
    \left\langle \mathcal{A}_{w}^{2}\right\rangle =\left(\frac{\partial^{2}Q_{\mathcal{A}_{w}}(p)}{\partial p^{2}}\right)_{p=0}=\left\langle \xi_{\beta}^{2}\right\rangle \int_{0}^{\infty}d\tau\varphi(\tau)\tau^{2+\beta},
    \label{awm2}
\end{eqnarray}

and the above integral is always divergent provided that $\beta>0$ and 
$0<\alpha<2$.
In consequence a careful analysis must be conducted in terms of the parameters $\alpha$ and $\beta$.  Whenever the moments are finite, the present approach also yields the correct prefactors for the expressions of $\left\langle \mathcal{A}_{w}\right\rangle $ and $\left\langle \mathcal{A}_{w}^{2}\right\rangle $; otherwise, only scaling exponents are meaningful. For those values for which the moments diverge we need to regularize the integrals by introducing a cutoff due to the observation time. This can be done in the integral over $\tau$ by integrating up to $t$. 
For $0<\alpha<1+\beta/2$ we introduce $t$ as the upper limit in the integral in \eqref{awm1}, so that
\begin{eqnarray}
\left\langle \mathcal{A}_{w}\right\rangle \sim\int_{t_{0}}^{t}\tau^{\frac{\beta}{2}-\alpha}d\tau\underset{t\gg t_{0}}{\sim}t^{1+\frac{\beta}{2}-\alpha}.
    \label{awm11}
\end{eqnarray}
For $1+\beta/2<\alpha<2$ the integral in \eqref{awm1} is convergent and
\begin{eqnarray}
    \left\langle \mathcal{A}_{w}\right\rangle =\alpha t_{0}^{\alpha}\left\langle \xi_{\beta}\right\rangle \int_{t_{0}}^{\infty}\tau^{\frac{\beta}{2}-\alpha}d\tau=\frac{\alpha\left\langle \xi_{\beta}\right\rangle t_{0}^{1+\frac{\beta}{2}}}{\alpha-1-\frac{\beta}{2}}.
    \label{awm12}
\end{eqnarray}
For the second moment the integral in \eqref{awm2} is always divergent, so that
\begin{eqnarray}
    \left\langle \mathcal{A}_{w}^{2}\right\rangle \sim\int_{t_{0}}^{t}\tau^{1+\beta-\alpha}d\tau\underset{t\gg t_{0}}{\sim}t^{2+\beta-\alpha}.
    \label{awm22}
\end{eqnarray}

On the other hand, the two first moments of the number of resets in $[0,t]$ appearing in Eqs. \eqref{man1} and \eqref{man2} are well known from the renewal theory. In the Laplace space \cite{GoLu01} they can be written in terms of the Laplace transform of the waiting times between resets as
$$
\left\langle \tilde{N}(s)\right\rangle =\frac{\tilde{\varphi}(s)}{s\left(1-\tilde{\varphi}(s)\right)},\quad\left\langle \tilde{N}(s)^{2}\right\rangle =\frac{\tilde{\varphi}(s)\left(1+\tilde{\varphi}(s)\right)}{s\left(1-\tilde{\varphi}(s)\right)^{2}}
$$
and considering \eqref{FI} in the long time limit, we find
\begin{eqnarray}
    \left\langle N(t)\right\rangle \underset{t\to\infty}{\approx}\left\{ \begin{array}{cc}
\frac{t^{\alpha}}{b_{\alpha}\Gamma(1+\alpha)}, & 0<\alpha<1\\
\frac{t}{\left\langle t_{R}\right\rangle }, & \alpha>1
\end{array}\right.,\quad\left\langle N(t)^{2}\right\rangle \underset{t\to\infty}{\approx}\left\{ \begin{array}{cc}
\frac{2t^{2\alpha}}{b_{\alpha}^{2}\Gamma(1+2\alpha)}, & 0<\alpha<1\\
\frac{t^{2}}{\left\langle t_{R}\right\rangle ^{2}}, & \alpha>1
\end{array}\right. .
\label{mN}
\end{eqnarray}
Inserting \eqref{awm11}, \eqref{awm12} and \eqref{mN} into \eqref{man1} we obtain the first moments of $\mathcal{A}_{N}(t)$ in the long time limit
\begin{eqnarray}
\left\langle \mathcal{A}_{N}(t)\right\rangle \underset{t\gg t_{0}}{\approx}\left\{ \begin{array}{cc}
\sim t^{1+\frac{\beta}{2}}, & 0<\alpha<1\\
\sim t^{2+\frac{\beta}{2}-\alpha}, & 1<\alpha<1+\frac{\beta}{2}\\
\frac{(\alpha-1)\left\langle \xi_{\beta}\right\rangle t_{0}^{\frac{\beta}{2}}}{\alpha-1-\frac{\beta}{2}}t, & 1+\frac{\beta}{2}<\alpha<2
\end{array}\right.
\label{an1s}
\end{eqnarray}
for the first moment. 
For the second moment we have to distinguish two different situations: subdiffusive motion ($0<\beta<1$) and superdiffusive motion ($1<\beta<2$). For $0<\beta<1$ inserting \eqref{awm22} and  \eqref{mN} into  \eqref{man2} 
\begin{eqnarray}
\left\langle \mathcal{A}_{N}(t)^{2}\right\rangle \underset{t\gg t_{0}}{\sim}\left\{ \begin{array}{cc}
t^{2+\beta}, & 0<\alpha<1\\
t^{3+\beta-\alpha}, & 1<\alpha<1+\beta\\
t^{2}, & 1+\beta<\alpha<2.
\end{array}\right.
\label{an2s1}
\end{eqnarray}
while for $1<\beta<2$
\begin{eqnarray}
\left\langle \mathcal{A}_{N}(t)^{2}\right\rangle \underset{t\gg t_{0}}{\sim}\left\{ \begin{array}{cc}
t^{2+\beta}, & 0<\alpha<1\\
t^{3+\beta-\alpha}, & 1<\alpha<2.\\
\end{array}\right.
\label{an2s2}
\end{eqnarray}
In figure 7 we plot the temporal evolution of the two first moments of the absolute area accumulated during the complete resetting events for a SBM. The agreement between the results of numerical simulations and the temporal scalings predicted by  Eqs. \eqref{an1s}, \eqref{an2s1} and \eqref{an2s2} is excellent.  

\begin{figure}[h!]
    \centering
    \includegraphics[width=0.75\linewidth]{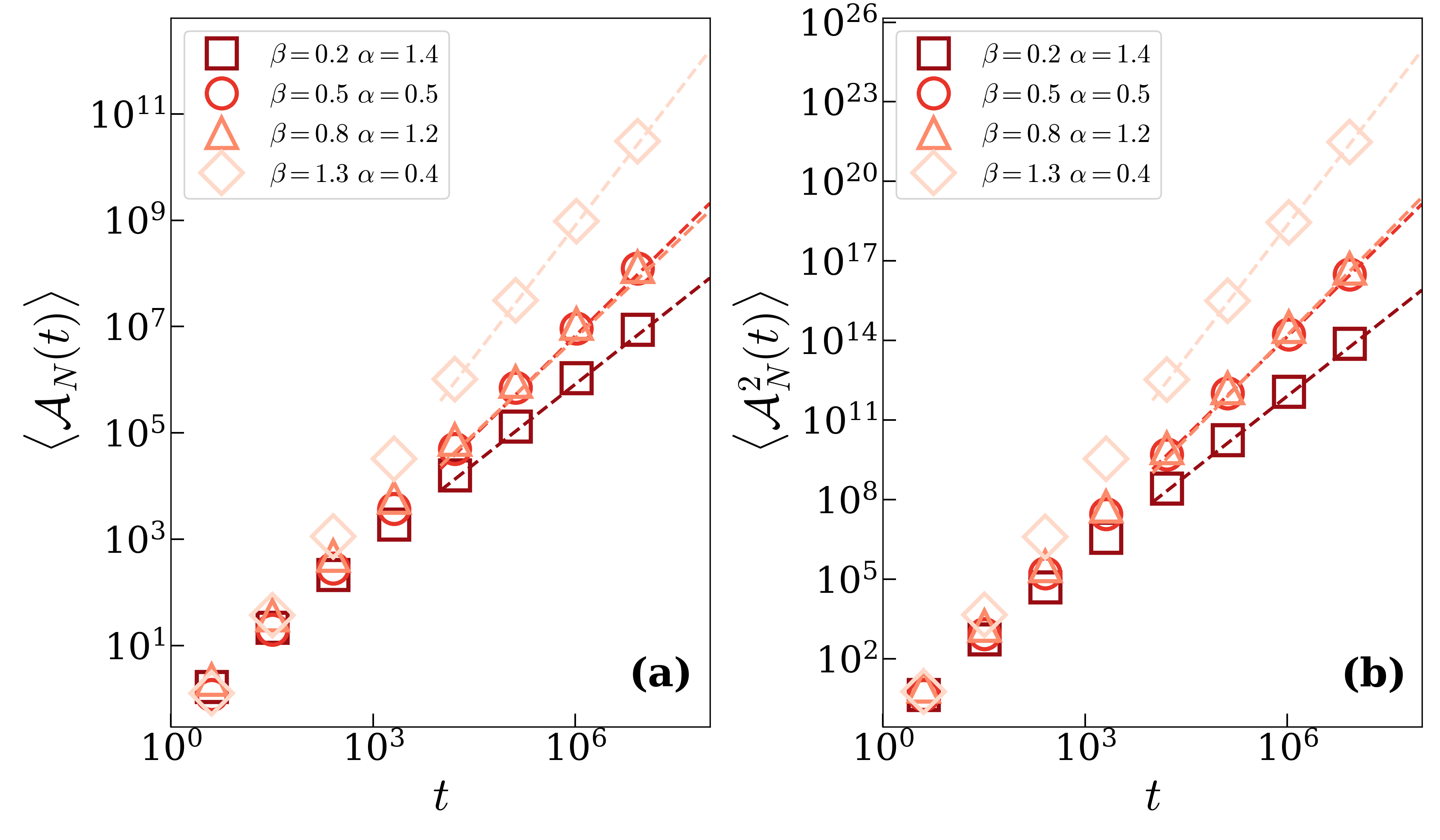}
    \caption{First (a) and second (b) moments of the waiting-time contribution $\mathcal{A_{N}}(t)$ under power-law resetting. Results are shown for different values of the resetting exponent $\alpha$ and the diffusion exponent $\beta$. Symbols correspond to stochastic simulations, while dotted lines denote the temporal scaling predictions in the long time limit given in Eqs. \eqref{an1s}, \eqref{an2s1} and \eqref{an2s2}.}
    \label{fig7}
\end{figure}
Finally we recall that the overall contribution to the absolute area under power-law resetting is the sum of the contribution from the waiting times plus the contribution of the backward time. According to \eqref{contrib}
 and \eqref{aa}
 $$
 \mathcal{A}_{r}(t)=\mathcal{A}_{N}(t)+\mathcal{A}_{\textrm{back}}(t).
 $$
In consequence, from \eqref{Absr}, \eqref{aab} and \eqref{an1s} the mean value of the absolute area below a SBM trajectory under power-law resetting is
\begin{eqnarray}
\left\langle \mathcal{A}_{r}(t)\right\rangle \underset{t\gg t_{0}}{\approx}\left\{ \begin{array}{cc}
\sim t^{1+\frac{\beta}{2}}, & 0<\alpha<1\\
\sim t^{2+\frac{\beta}{2}-\alpha}, & 1<\alpha<1+\frac{\beta}{2}\\
\frac{(\alpha-1)\left\langle \xi_{\beta}\right\rangle t_{0}^{\frac{\beta}{2}}}{\alpha-1-\frac{\beta}{2}}t, & 1+\frac{\beta}{2}<\alpha<2
\end{array}\right.
\end{eqnarray}
which is in agreement with \eqref{A1tf}.
Note that for $0<\alpha<1+\beta/2$ the contributions to the mean absolute area from the waiting times and from the backward time are of the same order while for $1+\beta/2<\alpha<2$ we have $ \left\langle \mathcal{A}_{N}(t)\right\rangle\gg  \left\langle \mathcal{A}_{\text{back}}(t)\right\rangle$ in the long time limit, so that in this case $\left\langle \mathcal{A}_{r}(t)\right\rangle \approx\left\langle \mathcal{A}_{N}(t)\right\rangle$. For $0<\beta<1$ the mean square value of the absolute area under a SBM under power-law resetting is, from \eqref{an2s1}, \eqref{abr2} and \eqref{ab2} given by
\begin{eqnarray}
    \left\langle \mathcal{A}_{r}(t)^{2}\right\rangle \underset{t\gg t_{0}}{\sim}\left\{ \begin{array}{cc}
t^{2+\beta}, & 0<\alpha<1\\
t^{3+\beta-\alpha}, & 1<\alpha<1+\beta\\
t^{2}, & 1+\beta<\alpha<2
\end{array}\right.
\end{eqnarray}
We note that for $0<\alpha<1+\beta$ the contributions to the second moment from the waiting times and from the backward time are of the same order while for $1+\beta<\alpha<2$ we have $ \left\langle \mathcal{A}_{N}(t)^2\right\rangle\gg  \left\langle \mathcal{A}_{\text{back}}(t)^2\right\rangle$. For $1<\beta<2$ we find from \eqref{an2s2}, \eqref{abr2} and \eqref{ab2}
\begin{eqnarray}
    \left\langle \mathcal{A}_{r}(t)^{2}\right\rangle \underset{t\gg t_{0}}{\sim}\left\{ \begin{array}{cc}
t^{2+\beta}, & 0<\alpha<1\\
t^{3+\beta-\alpha}, & 1<\alpha<2\\
\end{array}\right.
\end{eqnarray}
This result is in complete agreement with \eqref{A2tf2}

\subsection{The maximum contribution}
We want to study the statistical properties of the maximum value of the absolute area accumulated during the waiting periods between resets. Let us consider a trajectory from 0 to $t$ affected by $N$ resets with a sequence of waiting times $\{\tau_1,\cdots,\tau_N\}$. Define
$$
\mathcal{A}_{\max}=\max\{\mathcal{A}_{w}(\tau_1),\cdots,\mathcal{A}_{w}(\tau_N)\} .
$$
Given $N$ resets from 0 to $t$, the probability that $\mathcal{A}_{\max}$ is higher than $a$ is 
\begin{eqnarray*}
    \mbox{Prob}\left(\mathcal{A}_{\max}>a|N\right)&=&1-\mbox{Prob}\left(\mathcal{A}_{\max}<a|N\right)=1-\left[\int^{a}_{0}P_{\mathcal{A}_w}(a')da'\right]^{N}\\&=&1-\left[1-\int^{\infty}_{a}P_{\mathcal{A}_{w}}(a')da'\right]^{N}.
\end{eqnarray*}
Since $\mbox{Prob}\left(\mathcal{A}_{\max}>a|N\right)$ is the cumulative probability of $\mathcal{A}_{\max}$ given $N$ resets, the probability  that $\mathcal{A}_{\max}=a$ given that $N$ resets occur up to time $t$ is given by
\begin{eqnarray}
P_{\mathcal{A}_{\max}}(a|N)=-\frac{\partial}{\partial a}\mbox{Prob}\left(\mathcal{A}_{\max}>a|N\right)=NP_{\mathcal{A}_{w}}(a)z(a)^{N-1},   
\label{pamax1}
\end{eqnarray}
where we have defined
$$
z(a)=\int^{a}_{0}P_{\mathcal{A}_{w}}(a')da'.
$$
If we define $P_{ \mathcal{A}_{\max}} (a,N,t)$ as the joint probability of $\mathcal{A}_{\max}$ and $N$ in the observation time $t$, then $P_{ \mathcal{A}_{\max}} (a,N,t)=P_{ \mathcal{A}_{\max}} (a|N)P(N|t)$, so that
\begin{eqnarray}
       P_{\mathcal{A}_{\max}}(a,t)=\sum_{N=0}^{\infty}P_{\mathcal{A}_{\max}}(a,N,t)=\sum_{N=0}^{\infty}P_{\mathcal{A}_{\max}}(a|N)P(N|t)=P_{A_{w}}(a)\sum^{\infty}_{N=0}Nz(a)^{N-1}P(N|t), 
       \label{pamax2}.
\end{eqnarray}
Performing the Laplace transform of \eqref{pamax2} with respect to time and using \cite{GoLu01}
$$
P(N|s)=\frac{1-\varphi(s)}{s}\varphi(s)^{N}
$$
we can perform the sum over $N$ to find
\begin{eqnarray}
     P_{\mathcal{A}_{\max}}(a,s)=P_{A_{w}}(a)\frac{\varphi(s)^{-1}-1}{s\left[\varphi(s)^{-1}-z(a)\right]^{2}},
     \label{pamaxs}
\end{eqnarray}
which is the PDF of $\mathcal{A}_{\max}$. 
Introducing the long time limit approximation of the waiting time PDF given in Eq. \eqref{FI} into \eqref{pamaxs} we find
$$
P_{\mathcal{A}_{\max}}(a,s)\underset{s\to0}{\approx}P_{A_{w}}(a)\left\{ \begin{array}{cc}
\frac{b_{\alpha}s^{\alpha-1}}{\left(1-z(a)+b_{\alpha}s^{\alpha}\right)^{2}}, & 0<\alpha<1\\
\frac{\left\langle t_{R}\right\rangle }{\left(1-z(a)+\left\langle t_{R}\right\rangle s\right)^{2}}, & 1<\alpha<2
\end{array}\right.
$$
Now we perform the Laplace inversion but the cases $0<\alpha<1$ and $1<\alpha<2$ must be analyzed separately.

Consider first $0<\alpha<1$. In this case $P_{\mathcal{A}_{\max}}(a,s)$ can be rewritten as
$$
P_{\mathcal{A}_{\max}}(a,s)\underset{s\to0}{\approx}-\frac{1}{\alpha}P_{A_{w}}(a)\frac{\partial}{\partial s}\frac{1}{1-z(a)+b_{\alpha}s^{\alpha}},
$$
whose Laplace inversion can be expressed in terms of the two-parameter Mittag-Leffler function as
\begin{eqnarray}
       P_{\mathcal{A}_{\max}}(a,t)\underset{t\to\infty}{\approx}P_{A_{w}}(a)\frac{t^{\alpha}}{\alpha b_{\alpha}}E_{\alpha,\alpha}\left(-(1-z(a))\frac{t^{\alpha}}{b_{\alpha}}\right). 
       \label{pamax11}
\end{eqnarray}
For $1<\alpha<2$, after Laplace inversion we find
\begin{eqnarray}
  P_{\mathcal{A}_{\max}}(a,t)\underset{t\to\infty}{\approx}P_{A_{w}}(a)\frac{t}{\left\langle t_{R}\right\rangle }\exp\left[-(1-z(a))\frac{t}{\left\langle t_{R}\right\rangle }\right].  
  \label{pamax22}
\end{eqnarray}
To compute the two first moments of $\mathcal{A}_{\max}(t)$ we need to find how $z(a)$ depends on $a$. Inserting \eqref{Pa0} into the definition of $z(a)$ we obtain
$$
z(a)=\int^{\infty}_{0}d\xi_{\beta}P(\xi_{\beta})\int^{(a/\xi_{\beta})^{\frac{2}{2+\beta}}}_{0}\varphi(\tau)d\tau.
$$
Making use of the power-law expression \eqref{pl} for the waiting time PDF we finally find
\begin{eqnarray}
    z(a)=1-t^{\alpha}_{0}a^{-\frac{2\alpha}{2+\beta}}\int^{a/t^{1+\frac{\beta}{2}}_{0}}_{0}\xi^{\frac{2\alpha}{2+\beta}}_{\beta}P(\xi_{\beta})d\xi_{\beta}\simeq1-t^{\alpha}_{0}\left\langle \xi^{\frac{2\alpha}{2+\beta}}_{\beta}\right\rangle a^{-\frac{2\alpha}{2+\beta}}
    \label{zap}
\end{eqnarray}
for $a\gg t^{1+\frac{\beta}{2}}_{0}$.

Introducing \eqref{pawap} and \eqref{zap} into \eqref{pamax11} we finally obtain for $0<\alpha<1$
\begin{eqnarray}
    P_{\mathcal{A}_{\max}}(a,t)\underset{\underset{a\gg t_{0}^{1+\beta/2}}{t\to\infty}}{\approx}\frac{2}{(2+\beta)\Gamma(1-\alpha)}\frac{\left\langle \xi^{\frac{2\alpha}{2+\beta}}\right\rangle t^{\alpha}}{a^{1+\frac{2\alpha}{2+\beta}}}E_{\alpha,\alpha}\left(-\frac{\left\langle \xi^{\frac{2\alpha}{2+\beta}}\right\rangle t^{\alpha}}{\Gamma(1-\alpha)a^{\frac{2\alpha}{2+\beta}}}\right).
    \label{pdfmax1}
\end{eqnarray}
This expression suggests the scaling form
\begin{eqnarray}
t^{1+\frac{\beta}{2}}P_{\mathcal{A}_{\max}}(a,t)\underset{\underset{a\gg t^{1+\beta/2}_{0}}{t\gg t_{0}}}{\approx}f_{\alpha,\beta}\left(\frac{a}{t^{1+\frac{\beta}{2}}}\right)
\end{eqnarray}
where
$$
f_{\alpha,\beta}(\chi)=\frac{2}{2+\beta}\frac{\mathcal{C}}{\chi^{1+\frac{2\alpha}{2+\beta}}}E_{\alpha,\alpha}\left(-\frac{\mathcal{C}}{\chi^{\frac{2\alpha}{2+\beta}}}\right),\quad\mathcal{C}=\frac{\left\langle \xi^{\frac{2\alpha}{2+\beta}}\right\rangle }{\Gamma(1-\alpha)},\quad\chi=\frac{a}{t^{1+\frac{\beta}{2}}}.
$$
We can be even more specific and approximate the Mittag-Leffler function for small and large argument to get a simpler expression for the tail and the inner part of the PDF. For small argument $E_{\alpha,\alpha}(-z)\underset{z\to0}{\approx}1$ and for large argument $E_{\alpha,\alpha}(-z)\underset{z\to\infty}{\approx}z^{-2}\alpha/\Gamma(1-\alpha)$ \cite{Go20}. Hence, at the tail of $P_{\mathcal{A}_{\max}}(a,t)$ we have
\begin{eqnarray}
t^{1+\frac{\beta}{2}}P_{\mathcal{A}_{\max}}(a,t)\underset{\underset{a\gg t^{1+\beta/2}}{t\gg t_{0}}}{\approx}\frac{2\mathcal{C}}{2+\beta}\left(\frac{t^{1+\frac{\beta}{2}}}{a}\right)^{1+\frac{2\alpha}{2+\beta}}
\end{eqnarray}
and at the inner part we have
\begin{eqnarray}
t^{1+\frac{\beta}{2}}P_{\mathcal{A}_{\max}}(a,t)\underset{\underset{t_{0}^{1+\beta/2}\ll a\ll t^{1+\beta/2}}{t\gg t_{0}}}{\approx}\frac{2\alpha}{(2+\beta)\Gamma(1-\alpha)\mathcal{C}}\left(\frac{t^{1+\frac{\beta}{2}}}{a}\right)^{1-\frac{2\alpha}{2+\beta}}.
\end{eqnarray}

\begin{figure}[h!]
    \centering
    \includegraphics[width=0.75\linewidth]{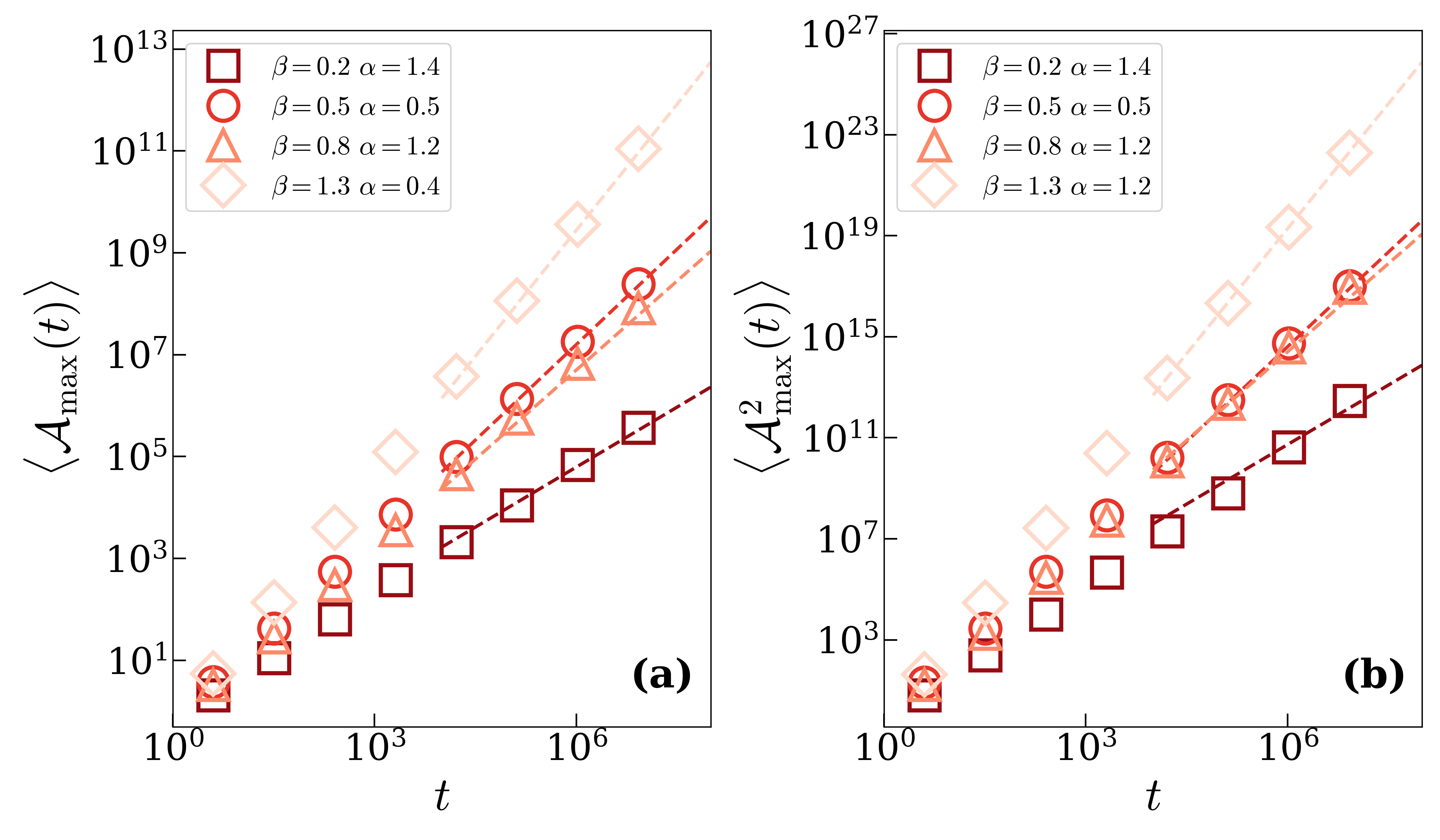}
    \caption{First (a) and second (b) moments of the maximum area contribution $\mathcal{A_{\max}}$ under power-law resetting. Results are shown for different combinations of the resetting exponent $\alpha$ and diffusion exponent $\beta$. Symbols correspond to stochastic simulations, while dotted lines denote the temporal scaling predictions in the long time limit given in Eqs. \eqref{amaxs1}, \eqref{amaxs2}, \eqref{amaxs3} and \eqref{amaxs4}.}
    \label{fig8}
\end{figure}

We make use of this scaling form to find the temporal scaling of the two first moments. The first moment is given by
\begin{eqnarray}
   \left\langle \mathcal{A}_{\max}(t)\right\rangle =\int^{\infty}_{0}aP_{\mathcal{A}_{\max}}(a,t)da\sim t^{1+\frac{\beta}{2}} 
   \label{amaxs1}
\end{eqnarray}
and the second moment follows analogously
\begin{eqnarray}
    \left\langle \mathcal{A}_{\max}(t)^{2}\right\rangle =\int^{\infty}_{0}a^{2}P_{\mathcal{A}_{\max}}(a,t)da\sim t^{2+\beta}.
    \label{amaxs2}
\end{eqnarray}
Let us now consider the case $1<\alpha<2$. Introducing \eqref{pawap} and \eqref{zap} into \eqref{pamax22} we obtain
\begin{eqnarray}
    P_{\mathcal{A}_{\max}}(a,t)\underset{\underset{a\gg t^{1+\beta/2}_{0}}{t\to\infty}}{\approx}\frac{2\alpha t^{\alpha}_{0}}{2+\beta}\frac{\left\langle \xi^{\frac{2\alpha}{2+\beta}}\right\rangle }{a^{1+\frac{2\alpha}{2+\beta}}}\frac{t}{\left\langle t_{R}\right\rangle }\exp\left(-\frac{\left\langle \xi^{\frac{2\alpha}{2+\beta}}\right\rangle t^{\alpha}_{0}}{a^{\frac{2\alpha}{2+\beta}}}\frac{t}{\left\langle t_{R}\right\rangle }\right),
    \label{pdfmax2}
\end{eqnarray}
which corresponds to the Fr\'echet PDF with exponent $\zeta=2\alpha/(2+\beta)$ and scale parameter $(\mathcal{D}t)^{1/\zeta}$, where $\mathcal{D}=\left\langle \xi^{\frac{2\alpha}{2+\beta}}\right\rangle t^{\alpha}_{0}/\left\langle t_{R}\right\rangle $. In consequence, the rescaled variable $a/(\mathcal{D}t)^{1/\zeta}$ in the long time limit converges asymptotically to a Fr\'echet random variable of index $\zeta$. This result is somewhat to be expected, given that $\tau_{\max}/t^{1/\alpha}$ also follows a Fr\'echet distribution for $\alpha>1$ \cite{GoMaSh15}. However, in our case the convergence holds for any underlying random walk with the appropriate index. Note that \eqref{pdfmax2} also admits a scaling form. If we define $\rho=a/t^{\frac{2+\beta}{2\alpha}} $ then
\begin{eqnarray}
    t^{\frac{2+\beta}{2\alpha}}P_{\mathcal{A}_{\max}}(a,t)\underset{\underset{a\gg t^{1+\beta/2}_{0}}{t\gg t_{0}}}{\approx}g_{\alpha,\beta}\left(\frac{a}{t^{\frac{2+\beta}{2\alpha}}}\right)
\end{eqnarray}
where
$$
g_{\alpha,\beta}(\rho)=\frac{2\alpha\mathcal{D}}{2+\beta}\frac{e^{-\mathcal{D}\rho^{-\frac{2\alpha}{2+\beta}}}}{\rho^{1+\frac{2\alpha}{2+\beta}}}.
$$

\begin{figure}[h!]
    \centering
    \includegraphics[width=0.75\linewidth]{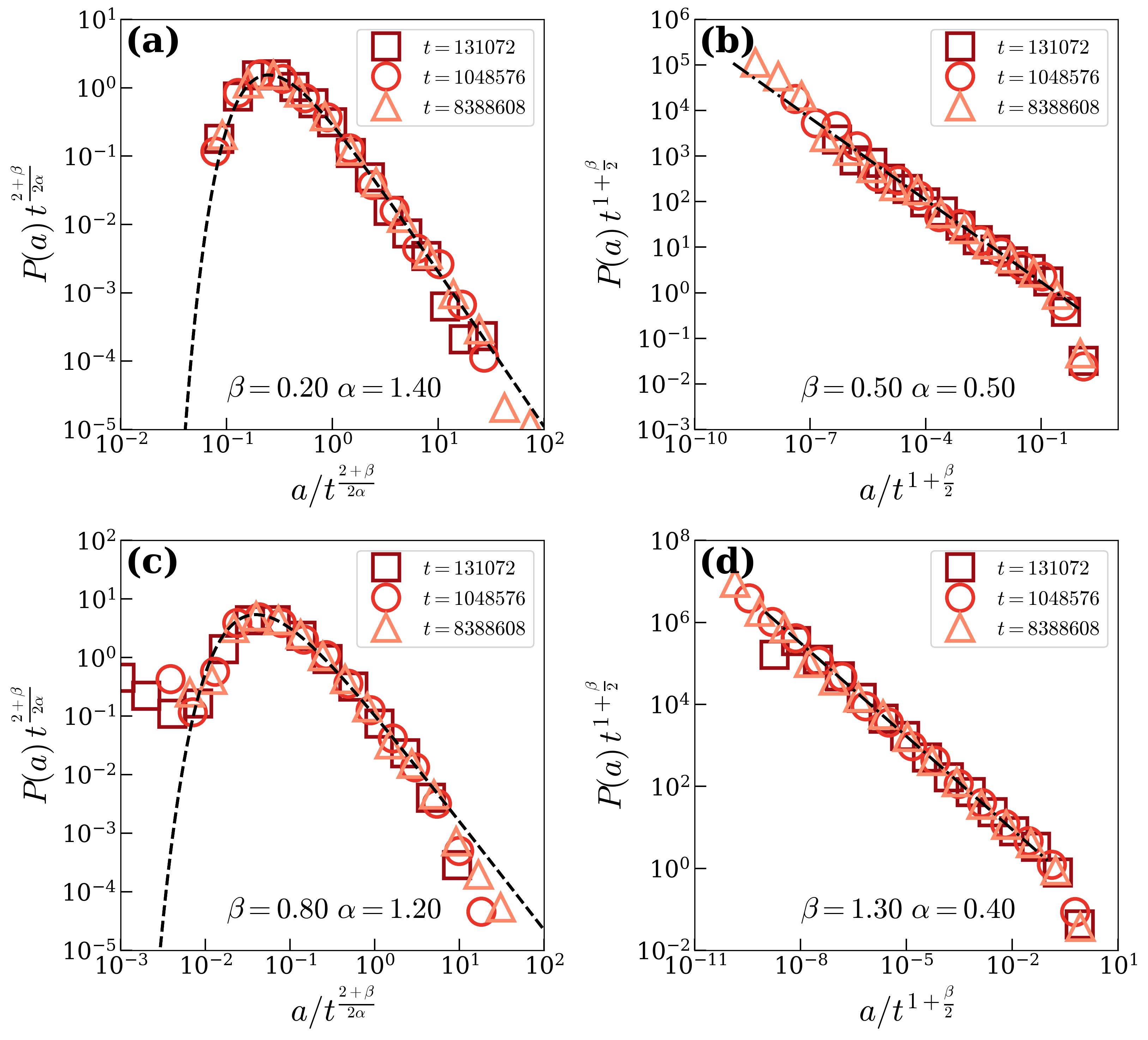}
    \caption{Scaling collapse of the probability density of the maximum area contribution $\mathcal{A}_{max}$ under power-law resetting. After rescaling according to the theoretical predictions, data obtained at different observation times collapse onto a single curve, confirming the asymptotic scaling form of $P(\mathcal{A}_{max})$.}
    \label{fig9}
\end{figure}

The temporal scaling of first moment can be found as
\begin{eqnarray}
   \left\langle \mathcal{A}_{\max}(t)\right\rangle =\int^{\infty}_{0}aP_{\mathcal{A}_{\max}}(a,t)da\sim t^{\frac{2+\beta}{2\alpha}} 
   \label{amaxs3}
\end{eqnarray}
and the second moment
\begin{eqnarray}
   \left\langle \mathcal{A}_{\max}(t)^{2}\right\rangle =\int^{\infty}_{0}a^{2}P_{\mathcal{A}_{\max}}(a,t)da\sim t^{\frac{2+\beta}{\alpha}} .
   \label{amaxs4}
\end{eqnarray}
It is interesting to note that the above scalings can be also derived from the statistical properties of the maximum waiting times. If $\tau_{\max}=\max\{\tau_1,\cdots,\tau_N\}$ then $\mathcal{A}_{\max}=\tau_{\max}^{1+\frac{\beta}{2}}\xi_\beta$, so that $\left\langle \mathcal{A}_{\max}\right\rangle =\left\langle \tau^{1+\frac{\beta}{2}}_{\max}\right\rangle \left\langle \xi_{\beta}\right\rangle $ and $\left\langle \mathcal{A}^{2}_{\max}\right\rangle =\left\langle \tau^{2+\beta}_{\max}\right\rangle \left\langle \xi^{2}_{\beta}\right\rangle$. For $0<\alpha<1$ one has the scaling $\tau_{\max}\sim t$ while for $1<\alpha<2$ the scaling is $\tau_{\max}\sim t^{1/\alpha}$ \cite{GoMaSh15}. From these scalings one readily recovers the scalings \eqref{amaxs1}, \eqref{amaxs2}, \eqref{amaxs3}, \eqref{amaxs4}. 
In figure \ref{fig9} we plot the PDF of $\mathcal{A}_{\max}(t)$ in terms of the scaling variable.  

In figure \ref{fig8} we plot the temporal evolution of the two first moments of the absolute area accumulated during the longest waiting periods of complete resetting events for a SBM. The agreement between the results of numerical simulations and the temporal scalings predicted by  Eqs. \eqref{amaxs1}, \eqref{amaxs2} \eqref{amaxs3} and \eqref{amaxs4} is excellent.  

In figure \ref{fig9} we plot the PDFs of $\mathcal{A}_{max}(t)$ for the different regimes found in our theoretical predictions. The scaling collapse is confirmed by numerical simulations and the agreement with our theoretical results is also excellent.   

It is worthy to note that if we compare the temporal scalings found for the two first moments of $\mathcal{A}_{\max}(t)$ in Eqs. \eqref{amaxs1}, \eqref{amaxs2}, \eqref{amaxs3} and \eqref{amaxs4} with the temporal scalings of the two first moments of $\mathcal{A}_{N}(t)$ in Eqs. \eqref{an1s}, \eqref{an2s1} and \eqref{an2s2} one can see that only for $0<\alpha<1$ the contribution to the absolute area from the sum of the accumulated contributions during the waiting periods between resets is of the same order that the contribution from the longest waiting period. This means that in this case the contribution to the area from the longest period dominates the statistics of $\mathcal{A}_{N}(t)$.

 \section{Discussion}
\label{sec:discussion}

The statistical properties of the absolute area under SBM with stochastic resetting reveals a rich variety of physical regimes. This behavior is governed by a delicate competition between the anomalous exponent $\beta \in (0, 2)$ and the tail of the waiting time PDF between resets. To elucidate the microscopic mechanisms driving ergodic and non-ergodic dynamics, we have decomposed the total area $\mathcal{A}_r$ into three fundamental components: the cumulative area over completed waiting periods between resets $\mathcal{A}_N$, the area generated during the backward recurrence time $\mathcal{A}_{\text{back}}$, and the maximum single-interval area contribution $\mathcal{A}_{\text{max}}$. A comprehensive overview of the asymptotic moment scaling, ergodicity breaking parameters, and dominant mechanisms across all parameter ranges is summarized in Table~\ref{tab:results}.

\begin{table*}[t] 
\caption{\label{tab:results} Summary of the temporal scalings and dominant mechanisms for the absolute area under SBM with stochastic resetting. } 
\small 
\begin{tabular}{llllll} \hline\hline \textbf{Resetting} & \textbf{$\beta$} & \textbf{$\langle \mathcal{A}_r(t)\rangle$} & \textbf{$\langle \mathcal{A}_r(t)^2\rangle$} & \textbf{EB} & \textbf{Dominant mechanism} \\ \hline Poissonian & All $\beta\in(0,2)$ & $\sim t$ & $\sim t^2$ & EB $=0$ &  \\ \hline Power-law \\  $0<\alpha<1$ & All $\beta\in(0,2)$  & $\sim t^{1+\beta/2}$ & $\sim t^{2+\beta}$ & const. & $\mathcal A_N\sim\mathcal A_{\rm back}\sim\mathcal A_{\max}$\\ \hline Power-law \\  $1<\alpha<2$ & Subdiffusive, $1<\alpha<1+\beta/2$ & $\sim t^{2+\beta/2-\alpha}$ & $\sim t^{3+\beta-\alpha}$ & $\sim t^{\alpha-1}$ & $\mathcal A_N\sim\mathcal A_{\rm back}\gg\mathcal A_{\max}$. \\  & Subdiffusive, $1+\beta/2<\alpha<1+\beta$ & $\sim t$ & $\sim t^{3+\beta-\alpha}$ & $\sim t^{1+\beta-\alpha}$ & $\langle\mathcal A_N\rangle\gg \langle\mathcal A_{\rm back}\rangle$ \\ & & & & & $\left\langle \mathcal{A}^{2}_{N}\right\rangle \sim\left\langle \mathcal{A}^{2}_{\text{back}}\right\rangle $ \\  & Subdiffusive, $1+\beta<\alpha<2$ & $\sim t$ & $\sim t^2$ & 0 & $\mathcal{A}_N\gg\mathcal{A}_{\text{back}}\gg \mathcal{A}_{\max} $ \\ \hline  & Superdiffusive, $1<\alpha<1+\beta/2$ & $\sim t^{2+\beta/2-\alpha}$ & $\sim t^{3+\beta-\alpha}$ & $\sim t^{\alpha-1}$ & $\mathcal A_N\sim\mathcal A_{\rm back}\gg\mathcal A_{\max}$. \\  & Superdiffusive, $1+\beta/2<\alpha<2$ & $\sim t$ & $\sim t^{3+\beta-\alpha}$ & $\sim t^{1+\beta-\alpha}$ & $\langle\mathcal A_N\rangle\gg \langle\mathcal A_{\rm back}\rangle$ \\ & & & & & $\left\langle \mathcal{A}^{2}_{N}\right\rangle \sim\left\langle \mathcal{A}^{2}_{\text{back}}\right\rangle $ \\ \hline\hline \end{tabular} \end{table*}

When the particle is reset at a constant rate $r$, the process erases its spatial memory at a finite characteristic scale $\langle \tau \rangle = 1/r$. As a consequence, the backward recurrence time contribution $A_{\text{back}}$ quickly saturates to a constant value at long times. For any exponent $\beta \in (0, 2)$ the EB parameter is zero. This confirms that Poissonian resetting acts as a robust stabilizer that suppresses trajectory-to-trajectory fluctuations and restores ergodicity regardless of whether the underlying motion is subdiffusive, standard, or superdiffusive. The PDF of the area converges to a Dirac delta function centered at the mean in the long time limit.
	
When the waiting times between resets follow a heavy-tailed power-law distribution with exponent $0 < \alpha < 2$, the existence of rare long waiting periods alters dramatically the area statistics. The system splits into distinct dynamical regimes depending on whether the mean resetting time diverges ($\alpha < 1$) or remains finite ($\alpha > 1$), as well as on the exponent $\beta$.

In the regime $0 < \alpha < 1$, the mean waiting time between resets diverges and the temporal heavy tail dominates the dynamics so strongly that the spatial differences between subdiffusion and superdiffusion become secondary. The total area is partitioned equally between the waiting times and the backward recurrence time. Crucially, the sum over the waiting times $\mathcal{A}_N$ is entirely dominated by the single largest waiting interval $\mathcal{A}_{\text{max}}$. This behavior exemplifies the \textit{Big Jump Principle}, where extreme individual events dictate the cumulative functional. As a result, trajectory-to-trajectory fluctuations do not decay, and the system exhibits Weak Ergodicity Breaking, characterized by a constant, positive value of EB.

When $1 < \alpha < 2$, the subdiffusive regime ($\beta < 1$) undergoes two distinct transitions, giving rise to three subregimes. In the first subregime ($1 < \alpha < 1 + \beta/2$), both $\mathcal{A}_N$ and $\mathcal{A}_{\text{back}}$ are of the same order but $\mathcal{A}_{\text{max}}$ follows Fréchet extreme value statistics, growing strictly slower than $\mathcal{A}_N$ and thus losing its dominance over the sum. Fluctuations diverge and the global area is highly nonergodic. 
Upon crossing into the second subregime ($1 + \beta/2 < \alpha < 1 + \beta$), $\langle \mathcal{A}_N\rangle$ dominates over the contribution from the backward time  $\langle \mathcal{A}_{\text{back}}\rangle$. However, the fluctuations are of the same order  $\langle \mathcal{A}_N(t)^2 \rangle \sim \langle \mathcal{A}_{\text{back}}(t)^2 \rangle$. This leads to divergent trajectory fluctuations, i,e., EB grows with time in the long time limit and the
global area is highly nonergodic. 
Finally, in the third subregime ($1 + \beta < \alpha < 2$), the system achieves a complete restoration of ergodicity. In this phase, $\mathcal{A}_N(t)$ dominates both the first and the second moments. The underlying physical mechanism relies on the spatial slowness of subdiffusion ($\beta < 1$), which prevents the particle for long excursions during long waiting times, effectively bounding area fluctuations.
	
In  contrast to subdiffusion, superdiffusive motion ($\beta > 1$) prevents the restoration of ergodicity across the entire interval $1 < \alpha < 2$. The rapid spatial spreading destroys the stabilizing effect of power-law resetting. The superdiffusive regime is partitioned into two subregimes. For $1 < \alpha < 1 + \beta/2$, the behavior mirrors the subdiffusive case with $1<\alpha<1+\beta/2$. For $1 + \beta/2 < \alpha < 2$,  $\mathcal{A}_N(t)$ dominates over $\mathcal{A}_{\text{back}}(t)$. Nevertheless, the ergodicity breaking parameter diverges in the long time limit. This case also mirrors the subdiffusive case with $1+\beta/2<\alpha<1+\beta$. The power-law resetting is fundamentally incapable of restoring ergodicity for superdiffusive motion.

\section{Conclusions}
\label{sec:conclusions}

In this work, we have developed a renewal framework  to analyze the statistical properties and ergodic behavior of the  absolute area  under SBM with anomalous diffusion exponent $\beta \in (0, 2)$ subject to stochastic resetting. By decomposing the total functional into the sum over the waiting times between resets $\mathcal{A}_N(t)$ and the backward recurrence time contribution $\mathcal{A}_{\text{back}}(t)$, we unveil  the boundaries separating ergodic and non-ergodic regimes. When waiting time are generated by a Poissonian process, the fluctuations of the absolute area are suppressed. As a result,  the ergodicity breaking parameter is zero for any $\beta \in (0, 2)$. In this domain, the backward time contribution saturates to a constant value, leaving $\mathcal{A}_N(t)$ as the sole dominant contribution to the area.

In contrast, heavy-tailed power-law resetting introduce a rich spectrum of non-ergodic behaviors governed by the interplay between rare resetting events and anomalous diffusion. For $0 < \alpha < 1$, the divergence of the mean waiting time causes Weak Ergodicity Breaking, characterized by a constant, positive value of EB, regardless of $\beta$. In this regime  the \textit{Big Jump Principle} holds, wherein the maximum contribution to the area, $\mathcal{A}_{\text{max}}$ dominates the entire sum over the waiting times $\mathcal{A}_N(t)$. When the mean waiting time is finite but its variance diverges ($1 < \alpha < 2$), the maximum contribution to the area does not dominate the sum over the resets. Crucially, our analysis uncovers an ergodic transition at $\alpha =1+\beta$ exclusive to subdiffusive motion: the slowness of subdiffusion keeps area fluctuations bounded. Normal ($\beta = 1$) and superdiffusive ($\beta > 1$) motions prevent the restoration of ergodicity across the entire range $1 < \alpha < 2$.
Our theoretical results may be useful in applications as cumulative spatial exposure, energy expenditure in intermittent search strategies, or residence times in crowded cellular environments govern system dynamics. Natural extensions of this work include generalizing the renewal framework to higher-dimensional geometries, spatial- or state-dependent resetting, or analyzing related stochastic functionals such as local times under broad resetting distributions.

\appendix

\section{Simulation Details}
In this Appendix, we describe the numerical algorithms employed to obtain the simulation results presented in the main text. All numerical data are generated from Monte Carlo simulations of ensembles comprising by $N$ independent particle trajectories.

The trajectories are generated using the Euler-Maruyama discretization scheme.  Within this, a single step of the stochastic trajectory is characterized by a time increment $\tau_i$ and a spatial increment $\Delta_i$. We fix $\tau_i = dt$ and define
\begin{equation}
\Delta_i = \sqrt{2D\,dt}\,\xi,
\end{equation}
where $\xi$ is a Gaussian random variable drawn from a normal distribution $\mathcal{N}(0,1)$. The diffusion coefficient is explicitly time-dependent and given by Eq. \eqref{Dsbm}. 

The model parameters were fixed as follows. For the SBM, we used $dt = 0.1$, $K = 1$, and $N =4 \times 10^{3}$. 

The quantities $\mathcal{A}(t)$, $\mathcal{A}_{\text{back}}(t)$, $\mathcal{A}_{N}(t)$ and $\mathcal{A}_{\max}(t)$ are computed from the corresponding discrete increments. These observables are evaluated at times that are integer multiples of the simulation time step. The probability density functions (PDFs) are constructed from the values of $\mathcal{A}_{\text{back}}(t)$ and $\mathcal{A}_{\max}(t)$ obtained in this manner. The binning procedure is chosen such that each bin contains sufficient statistics to enable a meaningful comparison with the corresponding continuous distributions.

\bibliography{main}
\end{document}